\documentclass[graybox]{svmult}

\usepackage{mathptmx}       % selects Times Roman as basic font
\usepackage{helvet}         % selects Helvetica as sans-serif font
\usepackage{courier}        % selects Courier as typewriter font
\usepackage{type1cm}        % activate if the above 3 fonts are
\usepackage{makeidx}         % allows index generation
\usepackage{graphicx}        % standard LaTeX graphics tool
\usepackage{multicol}        % used for the two-column index
\usepackage[bottom]{footmisc}% places footnotes at page bottom
\usepackage{amsmath,natbib}
\makeindex             % used for the subject index
\begin{document}

\title*{Companions of Stars: From Other Stars to Brown Dwarfs to Planets}
\subtitle{\large The Discovery of the First Methane Brown Dwarf}

% Use \titlerunning{Short Title} for an abbreviated version of
% your contribution title if the original one is too long
\author{Ben R. Oppenheimer}
% Use \authorrunning{Short Title} for an abbreviated version of
% your contribution title if the original one is too long
\institute{Ben R. Oppenheimer \at Department of Astrophysics, American Museum of Natural History, 79th Street at Central Park West, New York, NY 10025 USA, \email{bro@amnh.org}
}

\newcommand\aj{\rm{AJ}}%
          % Astronomical Journal
\newcommand\actaa{\rm{Acta Astron.}}%
  % Acta Astronomica
\newcommand\araa{\rm{ARA\&A}}%
          % Annual Review of Astron and Astrophys
\newcommand\apj{\rm{ApJ}}%
          % Astrophysical Journal
\newcommand\apjl{\rm{ApJ}}%
          % Astrophysical Journal, Letters
\newcommand\apjs{\rm{ApJS}}%
          % Astrophysical Journal, Supplement
\newcommand\ao{\rm{Appl.~Opt.}}%
          % Applied Optics
\newcommand\apss{\rm{Ap\&SS}}%
          % Astrophysics and Space Science
\newcommand\aap{\rm{A\&A}}%
          % Astronomy and Astrophysics
\newcommand\aapr{\rm{A\&A~Rev.}}%
          % Astronomy and Astrophysics Reviews
\newcommand\aaps{\rm{A\&AS}}%
          % Astronomy and Astrophysics, Supplement
\newcommand\azh{\rm{AZh}}%
          % Astronomicheskii Zhurnal
\newcommand\baas{\rm{BAAS}}%
          % Bulletin of the AAS
\newcommand\caa{\rm{Chinese Astron. Astrophys.}}%
  % Chinese Astronomy and Astrophysics
\newcommand\cjaa{\rm{Chinese J. Astron. Astrophys.}}%
  % Chinese Journal of Astronomy and Astrophysics
\newcommand\icarus{\rm{Icarus}}%
  % Icarus
\newcommand\jcap{\rm{J. Cosmology Astropart. Phys.}}%
  % Journal of Cosmology and Astroparticle Physics
\newcommand\jrasc{\rm{JRASC}}%
          % Journal of the RAS of Canada
\newcommand\memras{\rm{MmRAS}}%
          % Memoirs of the RAS
\newcommand\mnras{\rm{MNRAS}}%
          % Monthly Notices of the RAS
\newcommand\na{\rm{New A}}%
  % New Astronomy
\newcommand\nar{\rm{New A Rev.}}%
  % New Astronomy Review
\newcommand\pra{\rm{Phys.~Rev.~A}}%
          % Physical Review A: General Physics
\newcommand\prb{\rm{Phys.~Rev.~B}}%
          % Physical Review B: Solid State
\newcommand\prc{\rm{Phys.~Rev.~C}}%
          % Physical Review C
\newcommand\prd{\rm{Phys.~Rev.~D}}%
          % Physical Review D
\newcommand\pre{\rm{Phys.~Rev.~E}}%
          % Physical Review E
\newcommand\prl{\rm{Phys.~Rev.~Lett.}}%
          % Physical Review Letters
\newcommand\pasa{\rm{PASA}}%
  % Publications of the Astron. Soc. of Australia
\newcommand\pasp{\rm{PASP}}%
          % Publications of the ASP
\newcommand\pasj{\rm{PASJ}}%
          % Publications of the ASJ
\newcommand\qjras{\rm{QJRAS}}%
          % Quarterly Journal of the RAS
\newcommand\rmxaa{\rm{Rev. Mexicana Astron. Astrofis.}}%
  % Revista Mexicana de Astronomia y Astrofisica
\newcommand\skytel{\rm{S\&T}}%
          % Sky and Telescope
\newcommand\solphys{\rm{Sol.~Phys.}}%
          % Solar Physics
\newcommand\sovast{\rm{Soviet~Ast.}}%
          % Soviet Astronomy
\newcommand\ssr{\rm{Space~Sci.~Rev.}}%
          % Space Science Reviews
\newcommand\zap{\rm{ZAp}}%
          % Zeitschrift fuer Astrophysik
\newcommand\nat{\rm{Nature}}%
          % Nature
\newcommand\iaucirc{\rm{IAU~Circ.}}%
          % IAU Cirulars
\newcommand\aplett{\rm{Astrophys.~Lett.}}%
          % Astrophysics Letters and Communications
\newcommand\apspr{\rm{Astrophys.~Space~Phys.~Res.}}%
          % Astrophysics Space Physics Research
\newcommand\bain{\rm{Bull.~Astron.~Inst.~Netherlands}}%
          % Bulletin Astronomical Institute of the Netherlands
\newcommand\fcp{\rm{Fund.~Cosmic~Phys.}}%
          % Fundamental Cosmic Physics
\newcommand\gca{\rm{Geochim.~Cosmochim.~Acta}}%
          % Geochimica Cosmochimica Acta
\newcommand\grl{\rm{Geophys.~Res.~Lett.}}%
          % Geophysics Research Letters
\newcommand\jcp{\rm{J.~Chem.~Phys.}}%
          % Journal of Chemical Physics
\newcommand\jgr{\rm{J.~Geophys.~Res.}}%
          % Journal of Geophysical Research
\newcommand\jqsrt{\rm{J.~Quant.~Spec.~Radiat.~Transf.}}%
          % Journal of Quantitiative Spectroscopy and Radiative Trasfer
\newcommand\memsai{\rm{Mem.~Soc.~Astron.~Italiana}}%
          % Mem. Societa Astronomica Italiana
\newcommand\nphysa{\rm{Nucl.~Phys.~A}}%
          % Nuclear Physics A
\newcommand\physrep{\rm{Phys.~Rep.}}%
          % Physics Reports
\newcommand\physscr{\rm{Phys.~Scr}}%
          % Physica Scripta
\newcommand\planss{\rm{Planet.~Space~Sci.}}%
          % Planetary Space Science
\newcommand\procspie{\rm{Proc.~SPIE}}%
          % Proceedings of the SPIE
\let\astap=\aap
\let\apjlett=\apjl
\let\apjsupp=\apjs
\let\applopt=\ao

\maketitle

\abstract*{The discovery of the first methane brown dwarf provides a framework for describing the important advances in both fundamental physics and astrophysics that are due to the study of companions of stars.  I present a few highlights of the history of this subject along with details of the discovery of the brown dwarf Gliese 229B.  The nature of companions of stars is discussed with an attempt to avoid biases induced by anthropomorphic nomenclature.  With the newer types of remote reconnaissance of nearby stars and their systems of companions, an exciting and perhaps even more profound set of contributions to science is within reach in the near future.  This includes an exploration of the diversity of planets in the universe and perhaps soon the first solid evidence for biological activity outside our Solar System.}

\abstract{The discovery of the first methane brown dwarf provides a framework for describing the important advances in both fundamental physics and astrophysics that are due to the study of companions of stars.  I present a few highlights of the history of this subject along with details of the discovery of the brown dwarf Gliese 229B.  The nature of companions of stars is discussed with an attempt to avoid biases induced by anthropomorphic nomenclature.  With the newer types of remote reconnaissance of nearby stars and their systems of companions, an exciting and perhaps even more profound set of contributions to science is within reach in the near future.  This includes an exploration of the diversity of planets in the universe and perhaps soon the first solid evidence for biological activity outside our Solar System.}

\section{Why Objects Orbiting Stars are Important}
\label{Why}
The most obvious reason for looking for objects orbiting other stars is simply the possibility of finding a new type of object never before seen by human beings.  The extreme near vicinities of stars are generally inaccessible to observations because the star's light overwhelms significantly fainter companions or material.  

However, the regions of space around stars are now being revealed in unprecedented detail and a whole ``universe'' of diverse phenomena from different types of brown dwarfs, to planets and disks of material are being studied.  At the same time, some fundamental aspects of modern physics and astrophysics are due to the study of objects orbiting other stars.  A brief treatment of some of the history of studying companions of stars follows.

\subsection{An Early Use of Statistics in Science: The Discovery of Binary Stars}\label{stat}
John Michell, an English rector and friend of William Herschel, both of whom were keenly interested in establishing the distance scale to other stars (through parallax and other astrometric measurements), published a truly unique paper in 1767 in which he applied statistics to the distribution of positions of stars on the sky \citep{1767RSPT...57..234M}. Statistics was not even a formal mathematical field at the time.  Indeed, not until the early 1800s would Gauss publish the concept of what became known as a normal or ``Gaussian'' distribution.  Michell's paper is perhaps one of the earliest applications of statistics to modern science.\footnote{In this same paper, Michell derives the first reasonably accurate distance to Sirius, short by about a factor of 4, and discusses why stars twinkle.  Later Michell also proposed the concept of black holes based on Newtonian physics.  His early calculations are not vastly different from those that use general relativity.  Michell was truly a remarkable contributor to modern astrophysics.}

Michell simply took the angular distances between nearest neighbors of stars based on the various catalogs available at the time, and looked at their distribution, comparing that distribution to what one would expect if they were just randomly positioned in space.

He introduced this idea in the following manner:

\begin{quotation}
. . . from the apparent situation of the stars in the heavens, there is the highest probability that, either by the original act of the Creator or in consequence of some general law (such perhaps  as gravity), they are collected together in great numbers in some parts of Space while in others there are few or none.

The argument I intend to make use of, in order to prove this, is of that kind, which infers either design or some general law, from a general analogy, and the greatness of the odds against things having been in the present situation, if it was not owing to 
some such cause.

Let us then examine what it is probable would have been the least apparent distance of any two or more stars, any where in the whole heavens, upon the supposition that they had been scattered by mere chance, as it might happen.
\end{quotation}

Michell concludes that the stars must often be associated in clusters and in pairs through gravitational interaction, and that the probability that this statement is wrong is ``0.000076154.''

The importance of this cannot be overstated.  First there is the introduction of statistics into astronomy, but more importantly, this was the first concrete evidence that gravity actually operated outside the solar system.  Finally, it was the first statement that two stars might orbit each other, and that there exist in all likelihood gravitationally bound clusters of stars.  Incidentally, this was only his second publication in the field of astronomy.

William Herschel, slightly later, provided the final confirmation of the existence of binary stars, by being the first to measure astrometric reflex motion due to unseen companions of a number of the brightest stars.  This is not an easy task, because it requires the detection of aberrant motions of stars, i.e. motions that are not due to parallax or proper motion.\footnote{Proper motions of stars had been well established much earlier. In 1717 Edmund Halley used ancient catalogs, such as those of Ptolemy, Hipparchos and more recent positions of Aldeberan, Arcturus and Sirius to demonstrate motion \citep{1717RSPT...30..736H}.  Jacques Cassini, in 1740, confirmed the proper motion of Arcturus \citep{1740tads.book.....C}.  By 1770, Tobias Mayer's catalog of 80 proper motions established the direction of motion of the Sun with respect to the stars \citep{1770QB228.M46......}.}

He achieved this in a complex paper \citep{1803RSPT...93..339H} that details data covering a long period of observations:

\begin{quotation}
I shall therefore now proceed to give an account of a series of observations on double stars, comprehending a period of about 25 years, which, if I am not mistaken, will go to prove, that many of them are not merely double in appearance, but must be allowed to be real binary combinations of two stars, intimately held together by the bond of mutual attraction.
\end{quotation}

Thus, the discovery that stars have gravitationally bound companions resulted in evidence for a fundamental concept in modern physics: that gravity is a universal law.  The continued study of binary stars and faint companions of them led to even more interesting advances.

\subsection{Discovery of Degenerate Matter: The Companion of Sirius}\label{sirius}

In 1844 Friedrich Bessel examined decades of astrometry of the star Sirius.  He discovered an anomaly, and after considering various possible explanations, including perturbations in Earth's orbit about the Sun, concluded that there must be an unseen, but massive companion.  Its orbit suggested a 50 year period \citep{1844MNRAS...6R.136B}.

Alvan Clark, the great American telescope maker and observer himself, actually saw the companion of Sirius in 1862, as he recalls in his autobiography \citep{1889SidM....8..109C}.  Clark noted that it was about 8th magnitude or about 4000 times fainter than Sirius.  Using the orbit determined by Bessel, he concluded that the companion had to be about half the mass of Sirius due to its distance being twice that of Sirius from the center of gravity of their orbit.  That implied a mass about that of the Sun.  This was indeed a strange new type of object, but the physics of what it was would not begin to be revealed for another half century.

Walter Adams, a real pioneer in spectroscopic observations of stars, obtained a spectrum of ``Sirius B'' in 1915 \citep{1915PASP...27..236A}.  The spectrum was remarkable in that the temperature of the object would have to be around 25,000~K, while that of Sirius itself was only 10,000~K.  Given the fairly well known distance to the system, the measured luminosity, $L$, and this new temperature,  $T$, the Stefan-boltzmann relation $L=4\pi r^2 \sigma T^4$ gives a radius, $r$, roughly equal to that of the Earth.  With a mass of 1 M$_\odot$, this peculiar star had a density some 400,000 times that of the Earth.  At the same time as all of this was going on, the fundamentals of quantum mechanics were emerging.  Sirius B, now known to be the first example of a white dwarf, would later be explained through electron-degeneracy.  Work prompted by the existence of this white dwarf, the first observational constraints on degenerate matter, eventually led to Chandrasekhar's fundamental contributions to physics and stellar structure.\footnote{A fascinating story surrounds the companion of Sirius and a modern desire among some to believe that societies with ancient roots have known what modern science only recently discovered.  In 1950, an anthropologist studying the Dogon tribe in Mali, West Africa, published an account of a ceremony, supposedly performed since ``ancient'' times, which involves the fact that Sirius, central to their calendar system and mythology, is accompanied by an invisible, very heavy and metallic companion \citep{griaule1950}.  How could this ancient ceremony have been established unless an advanced, possibly extraterrestrial, civilization had visited thousands of years ago and told them about it?  However, as a number of scientists have pointed out \citep[e.g.]{1975Obs....95..215R}, the existence of Sirius B was known since the 1860s and its odd nature had been revealed by 1915 in America and Europe. Between 1915 and 1950 the Dogon had been visited by various European missionaries prior to Griaule's study.  Perhaps these visitors and their communications with the Dogon provide a far simpler explanation.}

\subsection{Further Motivation}
These are just a few examples of why studying companions of stars can provide major results in physics.  Others include the use of eclipsing binaries to measure stellar radii, and of course the first conclusive evidence of gravitational waves, through the detection of orbital period derivatives in binary pulsars.

However, perhaps the most important motivation for studying objects orbiting other stars is simply the fact that we live on a planet that orbits a star.  The parallel, that other stars may have planets, is no new idea.  Indeed, Epicurus c.~300BC wrote the following \citep{epicurus}:
\begin{quotation}
Moreover, there is an infinite number of worlds, some like this world, others unlike it. 
\end{quotation}

Some 1800 years later in 1584, Giordano Bruno \citep{bruno} made the daring assertion that 
\begin{quotation}
There are countless suns and countless earths all rotating around their suns in exactly the same way as the seven planets of our system.  We see only the suns because they are the largest bodies and are luminous, but their planets remain invisible to us because they are smaller and non-luminous. The countless worlds in the universe are no worse and no less inhabited than our Earth.
\end{quotation}
This statement is partly why Bruno was burned at the stake due to the dominance of the papacy, which seemed to have little tolerance for conjectures about the universe based on facts.  In addition, one can suppose that, even before Epicurus, people imagined that the stars were simply very distant versions of the Sun and that they might have their own planets going around them.  Note that Aristarchus advanced the idea of a heliocentric model of the solar system around the same time as Epicurus was alive \citep{North95}.\footnote{The actual book in which Aristarchus describes the first heliocentric model has not survived, but Archimedes in {\it The Sand Reckoner} leaves no doubt about Aristarchus's idea and the subsequent discussions it caused.  (See pp. 85-86 in \citet{North95}.)}

These historical facts, as well as a fascination with extremely high-resolution imaging and adaptive optics, are some of the motivations that led me to join Shri Kulkarni's group at Caltech for graduate school, in a project that was attempting to find any kind of very faint object orbiting nearby stars, using a relatively new technique.  The goal was to find brown dwarfs, objects that had been sought since Kumar's original technical report prepared when he worked at NASA's Goddard Institute for Space Studies in New York \citep{1962iss..rept....1K}.  However, I also saw an opportunity in developing the technique, in that advanced and refined versions of the technique would eventually lead to imaging and spectra of exoplanets.

\section{The Discovery of the First Methane Brown Dwarf}\label{sec:2}
Although the main subject of this section did not lead to a major revolution in physics, it is part of the history of an entirely new class of objects predicted by theory (see Kumar's Chapter in this volume).  Furthermore, the discovery of this unique object led to the first constraints on models of substellar atmospheres, and eventually, with the many other discoveries of so-called L and T-dwarfs, to a bridge between stellar astrophysics and planetary science---a bridge that is still being built as I write this.

In 1994, when I began my graduate research, brown dwarfs were the subject of numerous observational campaigns, primarily deep surveys of young star clusters.  Those sorts of searches and their successes are described in detail in Basri's and Rebolo's chapters in this volume.  Basri also details some later disproven claims of finding brown dwarf companions of stars.  One object, GD165B, at the time, was quite puzzling.  Orbiting a white dwarf, it seemed to be different from the coolest M-dwarfs, but it was never accepted as a bona fide brown dwarf \citep{1988Natur.336..656B}.  In retrospect, it is the first example of an early L-dwarf.  In addition, many brown dwarf searches were geared toward determining whether they could be a major component of baryonic dark matter.  Since this period, our understanding of brown dwarfs has completely changed.  Indeed they are a major part of the population of star-like objects, possibly almost as numerous as stars, but not a major contributor to dark matter.  Furthermore, there is a broad diversity in their emergent spectra, a diversity that is not a property of stars.  With more than 1000 known brown dwarfs, the field is rich and extremely active to this day.  Indeed, with possible connections to planets, and how the two classes of objects are related, brown dwarf research will remain important for the foreseeable future.

 What follows is partly a personal recollection of the events leading to the discovery of Gliese 229B and partly a description of the science.

\begin{figure}[ht]
\sidecaption[t]
\includegraphics[scale=.5]{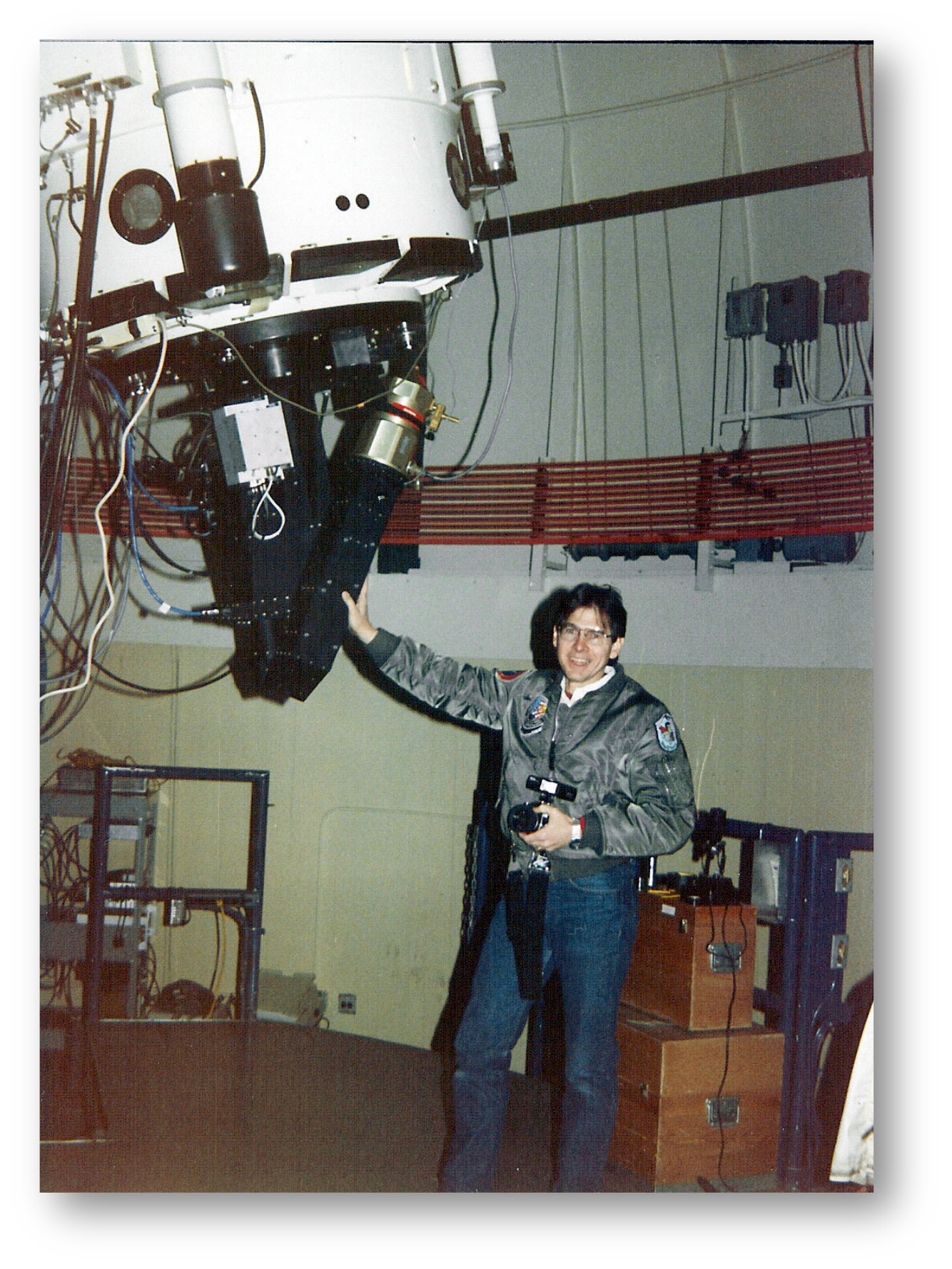}
\caption{Sam Durrance with the Johns Hopkins University Adaptive Optics Coronagraph in April 1989 at the 40-inch Swope Telescope at Las Campanas, Chile, prior to its move to the Palomar 60-inch telescope for the brown dwarf survey.  This instrument, one of the first stellar coronagraphs, was unique at the time.  It included a novel system that used the rejected starlight to control a fine guidance system using a quad-cell CCD and a piezo-electric tip/tilt mirror to keep the star centered on the occulting spot.  Plans existed to put a deformable mirror in the system, but funds were never raised to implement the full adaptive optics design of the instrument.}
\label{fig:AOC}       % Give a unique label
\end{figure}

\subsection{The Project at Palomar}
Shortly before my arrival at Caltech, Tadashi Nakajima, a post doctoral fellow working with Kulkarni, had forged a collaboration with a group based at Johns Hopkins University (JHU), led by Sam Durrance, and including David Golimowski.  Durrance and Golimowski, with Mark Clampin and Rob Barkhouser, had built a new type of coronagraph designed to image faint objects or disks around nearby stars.  This instrument (Fig.~\ref{fig:AOC}), called the JHU Adaptive Optics Coronagraph \citep[AOC]{1992ApOpt..31.4405G} had a high-bandwidth (100 Hz) image motion compensator, and the system was designed to have a deformable mirror put in it to permit higher-order wave front correction.  The higher-order system was never implemented, due to cost.

In 1989 the JHU team had deployed the AOC at the 40-inch Swope Telelscope at Las Campanas Observatory for a one year survey to find circumstellar disks and companions of nearby stars.  In December 1990 they moved the instrument to the 2.5-m duPont Telescope, the Carnegie Institute's premier telescope at the time.  This effort was a fantastic success, achieving some of the first images of the $\beta$ Pic disk in the optical that could be used for rigorous photometry within 100~AU of the star \citep{1993ApJ...411L..41G}, images of structure around young stars \citep[e.g.]{1993ApJ...410L..35C}, and analysis of close putative companions \citep{1993AJ....105.1108G}.  This body of work comprised a significant part of Golimowski's PhD thesis, completed in 1993 \citep{1994PhDT.........3G}.

By 1992 the instrument had been moved to the Palomar 60-inch Oscar Meyer Telescope, possible because the optical output of the telescope was quite similar to that of the duPont Telescope.  Nakajima, in close collaboration with Golimowski, started a campaign to image the remnant envelopes of star formation as seen in the optical, as reflection nebulae, around a slew of known pre-main-sequence stars \citep{1995AJ....109.1181N}.  In parallel, the concept of picking a set of stars with ages under 1 Gyr, but within 25 pc emerged at this point, to form the basis of a survey to find brown dwarfs around them.  The idea was simple.  Since brown dwarfs fade as they age, the survey should target the younger stars around which brown dwarfs would be brighter compared to their older counterparts  (assuming the star and its companion are coeval).  Of course, the brown dwarfs had to be there to begin with, and no brown dwarfs were known at the time.  This survey began just as I arrived in California \citep{1994ApJ...428..797N}.  

The general consensus was that the cooler these things were, the redder and redder they would be.  Indeed, a quick calculation, since the peak wavelength of a black body spectrum is in linear proportion to the temperature, suggested that a 1000~K object should have its peak emission wavelength around 6 times longer wavelength than that of the Sun (which is roughly 6000~K and peaks at about 0.6~$\mu$m), or roughly around 3 or 4~$\mu$m.\footnote{I visited Caltech in the spring of 1994 when I was deciding where to go to graduate school.  One of the best parts of that trip was that Shri invited me up to Palomar to spend a night at the Hale Telescope with him, Keith Matthews and Tom Hamilton.  I rented a ridiculously small car and went, not knowing that I was going to drive Shri and Tom back to Caltech.  Tom is rather tall and courageously squeezed into a half seat in the back of the car, his head tilted against the ceiling for the two hour drive.  As we drove down the mountain I had no idea Shri would test me.  One of the questions he asked was ``what is the peak wavelength of a 1000~K blackbody?'' Driving at 70 mph down the freeway, I recall suddenly becoming nervous, but also scrabbling my way through the answer.  I got it pretty close, but Shri then started spewing out all kinds of really simple quick relations of this sort, the kind you can do in your head in seconds.  I've never forgotten that, and though he did put me on the spot, I decided on that car ride that I really wanted to work with him.  It was a good choice.  Shri, though we had our differences at times, could not have been a better thesis advisor.}

The problem with the survey design, and the constant criticism the project came under, was that the coronagraph worked only in the optical band-passes.  The {\it z} band ($\sim 0.85$~$\mu$m) was as red as it could image, and the CCD's sensitivity at those wavelengths was somewhat diminished.  However, I think this is partly why the project was a success.  The general wisdom at the time was that to see a brown dwarf, instruments operating at infrared (IR) wavelengths were necessary, and, further, the longer the wavelength used, the better the chances of success were.  Numerous previous surveys used the assumption that cool brown dwarfs must be extremely red, and peak in the near IR.  In fact this was not the case.  T-dwarfs, as we now call them, peak near 1~$\mu$m and are actually blue in the IR, due to the complex thermochemistry of their atmospheres (cf. also Cushing's and Baraffe's chapters in this volume).  This is especially true because of the abundance of methane, which becomes stable below temperatures of about 1400~K.  Methane is a strong absorber in bands throughout the 1 to 5~$\mu$m range.  Incidentally, at the time, a number of papers suggested that these objects would be orders of magnitude different from black body spectra \citep[e.g.]{bhs93}, but only one group had predicted the huge effect of methane \citep{1995bmsb.conf...45T}.  In fact Takashi Tsuji had predicted the importance of methane only shortly after Kumar's first proposal of substellar objects \citep{1964AnTok...9.....T}.  Tsuji's work had not been incorporated into any survey designs prior to the discovery of the first methane brown dwarf.

When our survey started, though, we were not thinking of methane.  Rather, there was a strong indication from observations of the M-dwarfs, that anything cooler than them would not obey black body spectral characteristics, and we had hoped that these brown dwarfs would be bright enough to detect in the optical.  Regardless, the key difference between the Palomar AOC survey and others was that there was no color bias at all in the design of the observations and discovery criteria.  The only goal was to find any object around a nearby star that demonstrated that it was orbiting the star through detection of common proper motion.  This requirement for discovery, simply that the companion followed its presumed primary star through the sky, an extremely strong argument for physical association, had no bias in color.  Many other surveys, including surveys of star clusters, invoked color requirements for putative brown dwarf detection and follow-up \citep[e.g.][]{1995Natur.377..129R,1996ApJ...458..600B}

\subsection{Observing}
A month and a half after arriving in Pasadena, and in the midst of a heavy course load for my first year, I was up the mountain at Palomar for the first of what would be well over 230 nights at the observatory over the following four and a half years.  Tadashi was, very unfortunately, getting ill, so David Golimowski and I took over for a six night observing run in late October.  It is important to note here that, although very few results came out of the first year and a half of AOC observing, our primary goal on this observing trip was to confirm a putative companion to the star Gliese 105, a third or ``C'' component, which, according to the initial data from a previous observation with the AOC, would have to be one of the lowest mass stars ever found. With the long nights and hefty setup procedure for the AOC---we had to do almost everything ourselves, which was normal on the 60-inch---exhaustion was the norm.  In addition, Dave had to teach me everything, because he could not always be present at all the upcoming observing runs, and I was slated (and wanted) to conduct a majority of the observing for the survey.  

The AOC was a rather ``hands-on'' instrument.  Even to change filters, one had to run out of the control room, grab the new filter in its small holder, pull the old one out by unscrewing some fasteners, slide the new one in and run back as fast as possible to start a new exposure and not waste too much time with the camera shutter closed.

\begin{figure}[t]
\sidecaption[t]
\centerline{\includegraphics[scale=.47]{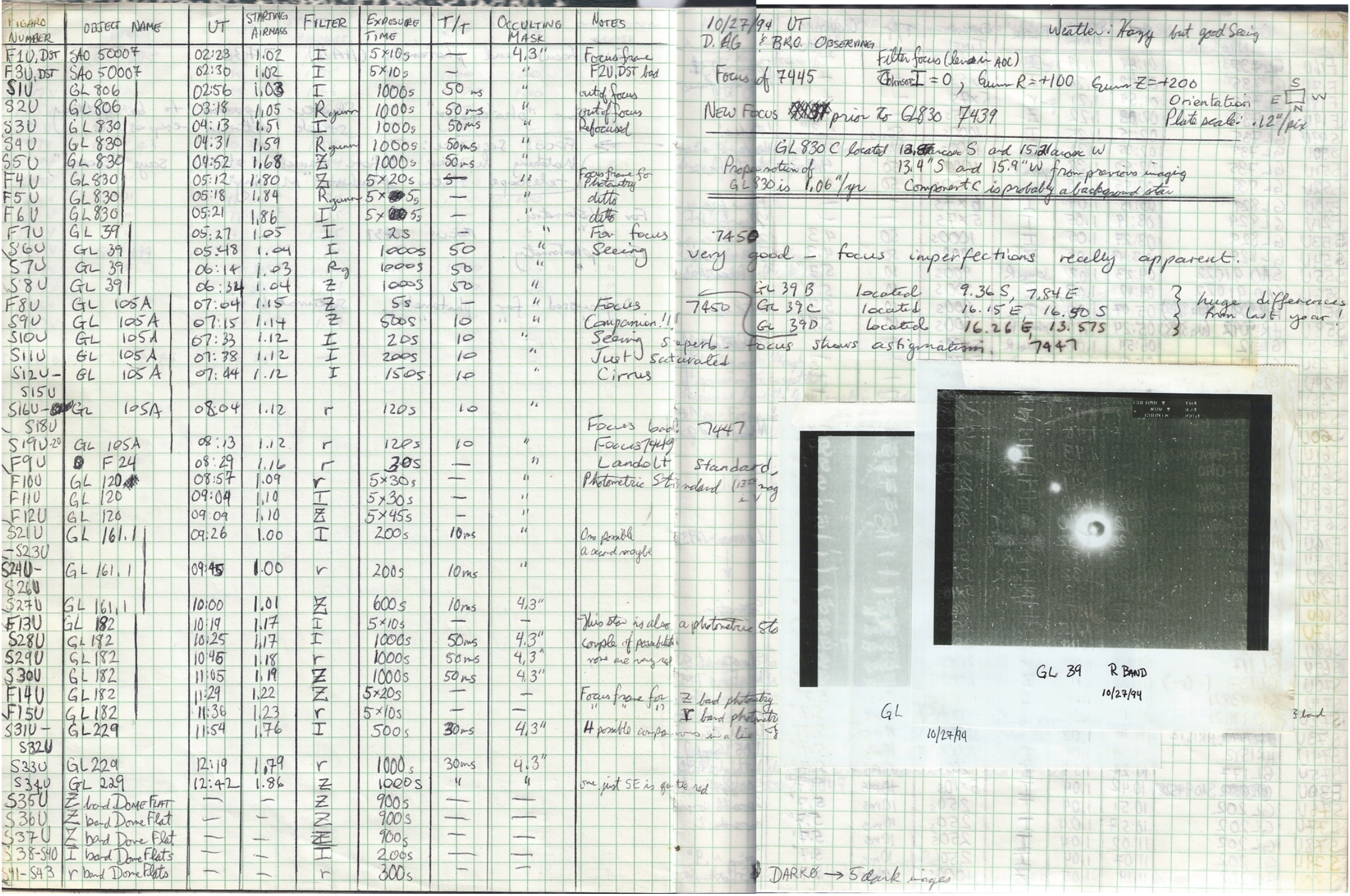}}
\caption{Observing log from the first night Gliese 229B was detected, October 27, 1994.  The bottom of the log shows that ``GL229'' was observed as the last star in the night, one with excellent seeing.  A short note for the last observation says ``one [possible companion] just SE is quite red.''  We were too tired to note anything else, and, in fact, had no idea what we had just observed.  The images taped into the log book were from an early video printer that allowed us to print images for reminders when reducing the data.}
\label{fig:log1}       % Give a unique label
\end{figure}

On 27 October 1994, Dave and I observed the star we called GL229.\footnote{After publication of the discovery of the companion to Gliese 229B, the IAU sent an admonishing letter to Shri that the designation GL was not appropriate because it conflicted with a U.~S.~Air Force catalog of celestial sources.  Officially, we had to call it Gliese 229.}  The night had excellent atmospheric conditions, especially for the Palomar 60-inch, with $I$-band measurements indicating roughly 0.5 to 0.4 arcsecond seeing.  Coronagraphic efficiency in removing starlight is an extremely strong function of the input image quality---partly why so many new coronagraphs have been built for adaptive optics systems since the 90s.  On this same night, we also observed Gliese 105C and confirmed its companionship \citep{1995ApJ...452L.125G,1995ApJ...444L.101G}.  This was a very exciting result from that night, but it would be upstaged quickly by events that were to unfold within 48 hours.

On that night, I was in charge of the list of target stars we were observing.  I guess because I was tired and excited about Gliese 105C, I forgot to mark the list to indicate that we had already observed Gliese 229.  I did note in the notebook (Fig.~\ref{fig:log1}), that there was something ``quite red'' to the south east of the primary star.  Note that the field of view of the instrument was about an arcminute, so nearly everything we observed had some kind of other star in the field of view.  This was slow work, and in any case, even if we found something interesting, we would have to wait until the following year to allow the star to complete a year's worth of proper motion before we could see whether a putative companion was actually traveling along with the star.  In any case, on this night, neither Dave nor I seemed to care about this little red dot next to Gliese 229.  

\begin{figure}[t]
\sidecaption[t]
\centerline{\includegraphics[scale=.47,angle=0]{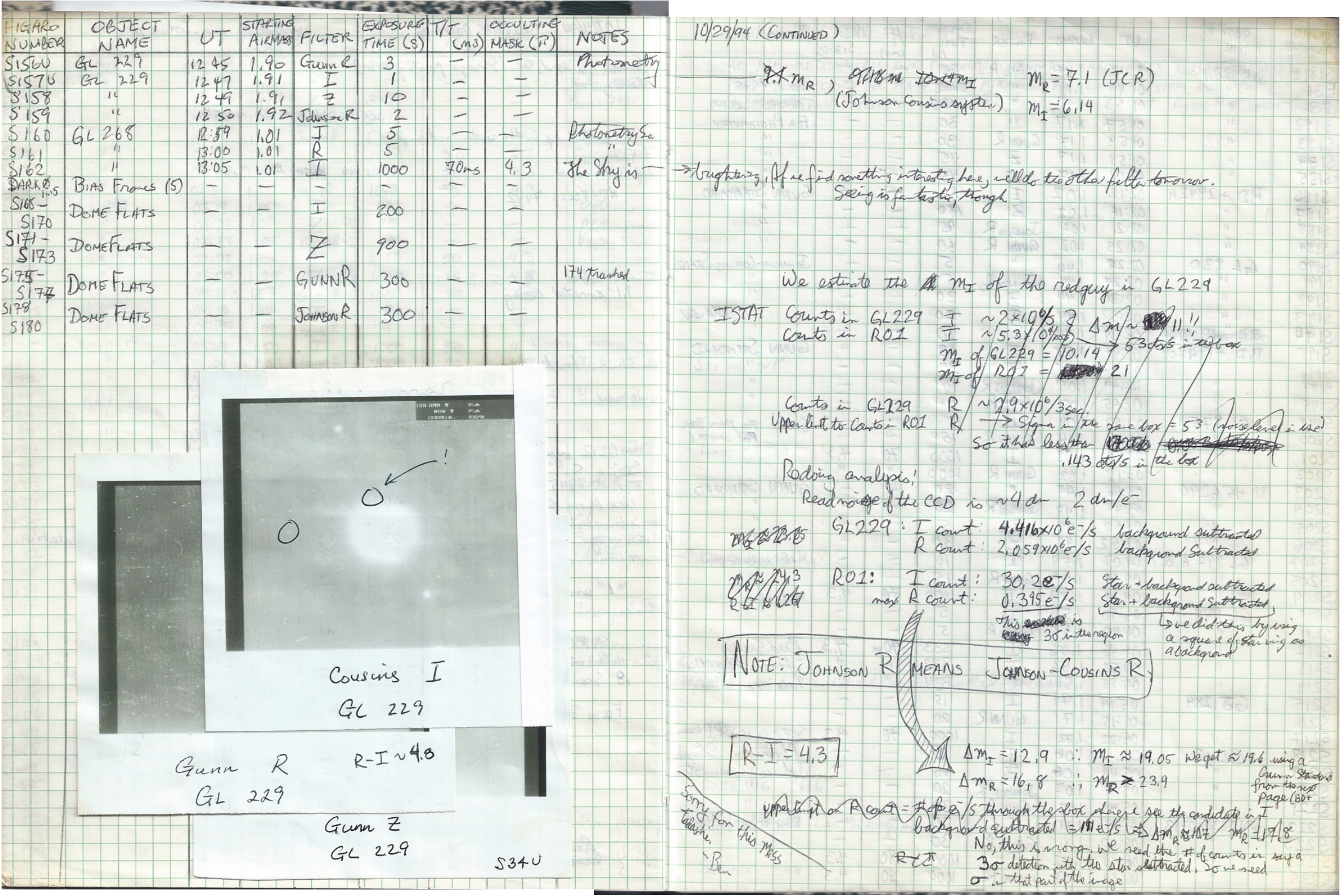}}
\caption{Observing log from the second night Gliese 229B was observed, October 29, 1994.  Again, ``GL229'' was observed near the end of the night, but this time around we realized this was a new type of object.  Scribbled at right are initial attempts to derive the magnitudes and colors of the companion.  None of this work could become public for another year, though, because we had to confirm that this point source, marked in the video printout at left, was in fact orbiting the star.}
\label{fig:log2}       % Give a unique label
\end{figure}

Two nights later, by virtue of the fact that I had failed to take ``GL229'' off our list of stars to observe, we pointed the telescope at it again.  This time, we knew we had something remarkable (Fig.~\ref{fig:log2}).  While we were taking darks and flat fields (roughly an hour or two of work after the Sun came up) Dave and I calculated and recalculated the photometry, in rough manner, since we did not have the data in the final format for analysis with a standard software package like IRAF.\footnote{At the time, many instruments produced data in non-standard formats that had to be converted at one's home institution where better computers were available.}  We realized that night that this object had an extremely red $R-I$ color of about 4.3$^{\rm m}$.  I was writing in Tadashi's notebook, which was generally just reserved for observing notes, so I added a small apology, "Sorry for this mess, Tadashi. ---Ben."  But Dave and I were very excited. 

I was so excited that, when we started observing the next night, I called Shri, who happened to be in Hawaii, using the relatively new Keck telescope with a beautiful spectrograph, called LRIS.    I told him what we had found and said he should get a spectrum of it immediately.  Shri, rightly so, did not.  I was a graduate student who had only been working with him for a month or so.  The Keck telescope was by far the largest in the world and not the kind of observatory that one would flippantly just point at some new object that someone is excited about because of rough calculations based on very raw data.  In retrospect, although I was disappointed Shri did not acquire an optical spectrum of Gliese 229B that night, he made the right call.  It turned out to be rather tricky, from an instrument configuration and data-analysis point of view, to obtain that spectrum later \citep{1998ApJ...502..932O}, and Shri probably would have wasted an hour of Keck time only to obtain data that was not useful.  The entire team would have to wait a year to confirm that Gliese 229 and this little red point source shared the same proper motion and indeed were orbiting each other.
 
\begin{figure}[t]
\sidecaption[t]
\centerline{\includegraphics[scale=.4,angle=0]{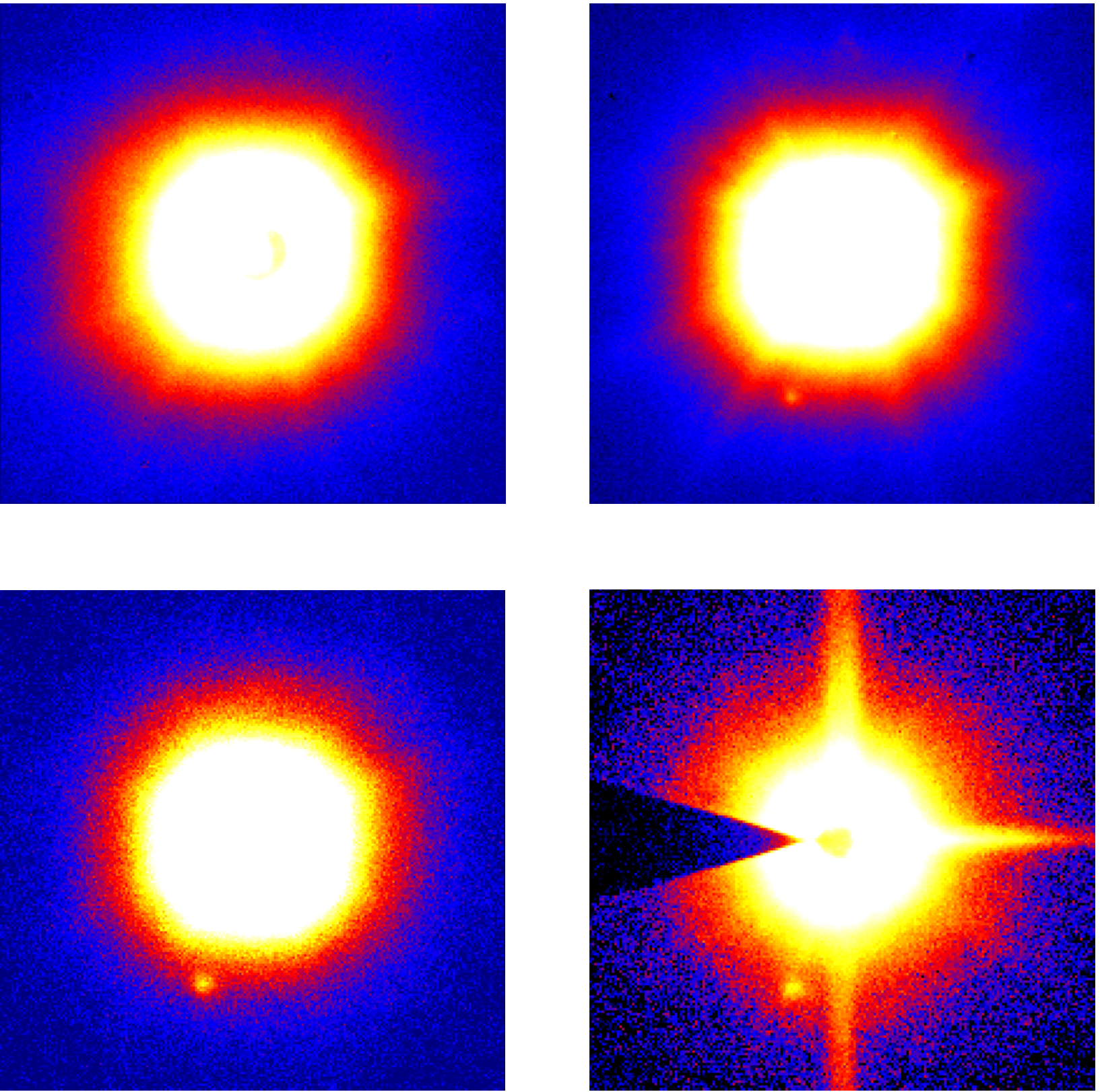} \includegraphics[scale=.26,angle=0]{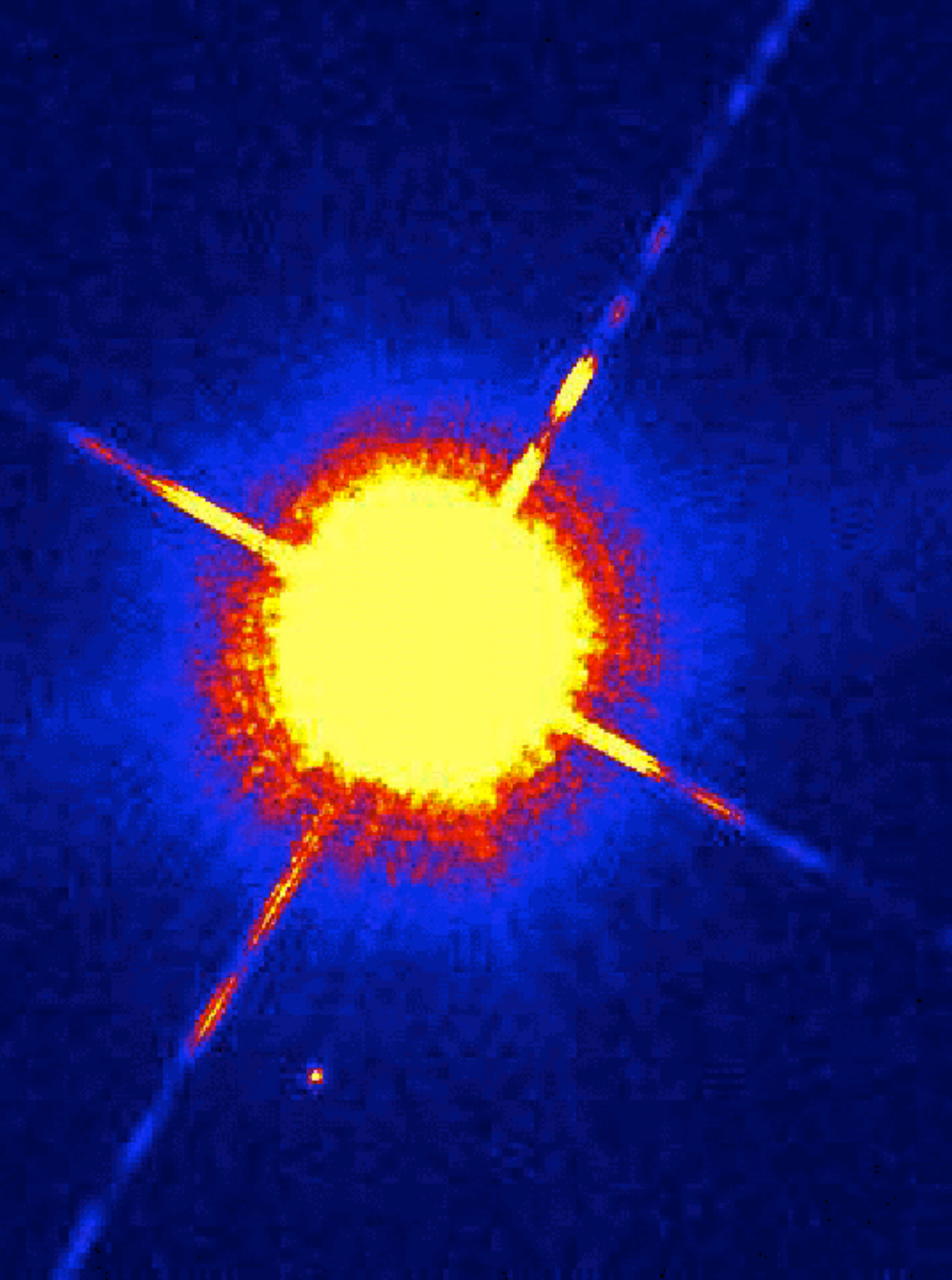}}
\caption{Discovery images of Gliese 229B.  Left: three optical ($r$, $i$, top and $z$, bottom left) and one $K$-band image (bottom right) from 1994 and 1995, adapted from \citet{1995Natur.378..463N}.  Right: Hubble Space Telescope image, demonstrating the enormous effect that image resolution has on revealing faint objects around bright stars.  This image is not taken with a coronagraph, while the others were.  Adapted from \citet{gbk98}. }
\label{fig:images}       % Give a unique label
\end{figure}

I am not a patient person, and this year of waiting to find out more and confirm this new object was a bit of agony.  I buried myself in my coursework, more observing and some other projects that I used to distract myself.  Everyone on the team was quiet.  We discussed nothing of the data, although after running a full analysis on it, the $r-z$ color was greater than 4.3$^{\rm m}$, a color no other star-like object shared.  During this time, I recall having to give a short talk on my work for my first year of graduate school, partly as a lead-in to the qualifying exams that, at that time, happened in the very beginning of the second-year.  I spoke to the department about the project, what we were doing and hoping to find.  It was a terrible talk, but I remember wanting to say something about this mysterious object.  I did not.  At the end of the talk, someone in the audience said, ``Brown dwarfs don't exist!  We have been looking for them for decades.  They aren't there.  You should work on something that exists!''  I did not respond to this.  Of course, he could have been right.  We did not know what we had found yet.

\begin{figure}[t]
\sidecaption[t]
\centerline{\includegraphics[scale=.4,angle=-90]{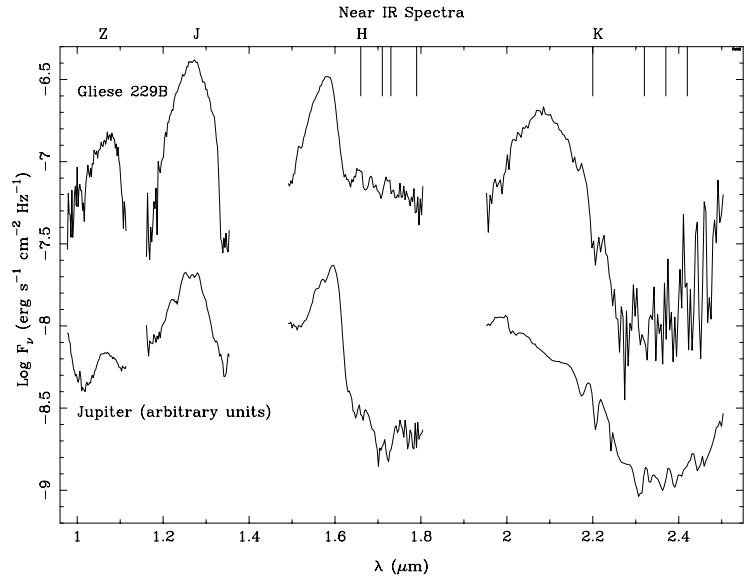}}
\caption{First spectrum of Gliese 229B compared with Jupiter, adapted from \citet{1995Sci...270.1478O}.  The locations of absorption band-heads due to methane is indicated by the vertical bars along the top of the plot in the $H$ and $K$-bandpasses.  }
\label{fig:spectrumScience}       % Give a unique label
\end{figure}

\subsection{``There's methane in that thing!''}
The eleven months between 29 October 1994 and 14 September 1995 passed, and Shri, Keith Matthews and I were back at Palomar, though on the 200-inch Hale telescope this time, with a wonderful instrument Keith built called D-78 (which simply stands for the 78$^{\rm th}$ cryogenic dewar in a series of pioneering infrared instruments, many of which Keith built).  D-78, an infrared camera, had a rudimentary fake coronagraph in it, which consisted of a metal finger that could be placed in the center of the field of view, to prevent saturation from a bright star.  Furthermore, it had the capability to take relatively low-resolution spectra of objects within the field of view, using a slit and a grism.  We obtained $JHK$ images and spectra of Gliese 229 and its companion.  Although we knew we had another observing run with the AOC on the 60-inch telescope only a week later, I had hoped that we could confirm that the companion and the star shared proper motion with this data.  Generally using two different instruments to conduct precision relative astrometry like this is not a good idea, but we observed a few calibration binary stars with well-known separations and thought we could try.  However, something more surprising happened.

\begin{figure}[t]
\sidecaption
\centerline{\includegraphics[scale=.4,angle=-90]{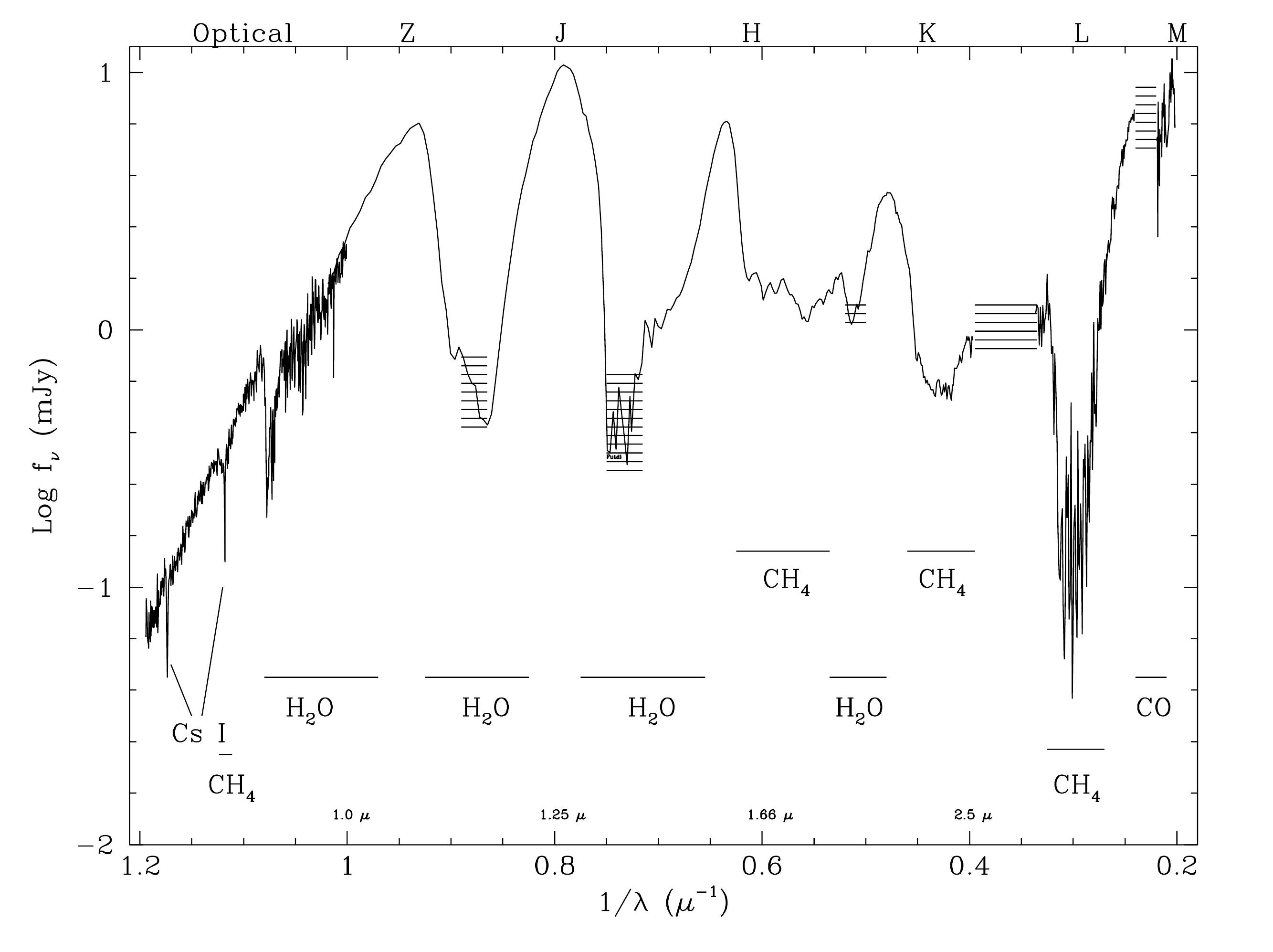}}
\caption{The full spectrum of Gliese 229B from optical to thermal IR bands \citep{1998ApJ...502..932O}.  Labeled horizontal bars indicate absorption bands due to methane (CH$_4$), water (H$_2$O) and carbon monoxide (CO).  Two atomic lines are also indicated and are due to neutral cesium (Cs {\small I}).}
\label{fig:spectrumFull}       % Give a unique label
\end{figure}

When Keith took the first grism spectrum (Fig.~\ref{fig:spectrumScience}), it showed up on the display and he said immediately (with some expletives removed), ``There's methane in that thing!''  He could see just from looking at the raw data that a large chunk of both the $H$ and $K$-bands were dark, right where methane is known to absorb light.  For reference, we observed Jupiter, which happened to be at a favorable observing angle, and which was the only real comparison we could make at the time.  Jupiter showed similar strong spectral features from methane absorption, as it had been known to exhibit for many years.  This companion of Gliese 229B was unique.  It had to be extremely cool to have methane in it, and there was no doubt at this point that it was either a companion of Gliese 229 or a foreground object that was extremely cool.   The astrometry demonstrating common proper motion, though necessary to establish companionship, was not needed to confirm that we had found something new.

I immediately began reducing the data at the telescope and Shri and I started writing the paper.  Within a week, we had generated the figure on the left side of Fig.~\ref{fig:images}, as well as Fig.~\ref{fig:spectrumScience}.  I also measured the astrometry between A and B using the D-78 data.  This work indicated that the two objects were indeed moving across the sky in the same direction and rate, although the astrometry was not as precise as I had hoped.  Hedging our bets, by 25 September 1995, we submitted the discovery paper, based on the astrometry and photometry obtained at that point.   On 3 October 1995, Shri and I obtained more AOC observations of the system and measured the astrometry (in between changing filters and taking more survey data).  The AOC astrometry was dead on.  Both the primary star and its companion were co-moving.  We sent an addendum to add this information into the paper submitted, and I submitted the paper on the spectrum (Fig.~\ref{fig:spectrumScience}) by 12 October 1995, four days after finishing the AOC observing trip.  These two papers, which we decided would be better as a pair rather than a single result, present the astrometric confirmation and photometric analysis of Gliese 229B \citep{1995Natur.378..463N} and its spectrum \citep{1995Sci...270.1478O}.  In the first paper, using the Stefan-Boltzmann relation, that Adams and others used for Sirius B (Sect.~\ref{sirius}), $L=4\pi r^2 \sigma T^4$, and estimating a radius based on brown dwarf models (roughly a Jupter radius, R$_{\rm J}$), we determined that the companion had a temperature of about 1000~K (which we compared with the models of \citet{1995bmsb.conf...45T}, for additional support), a luminosity of $4 \times 10^{-6}$ L$_\odot$, and a probable mass between 20 and 50 Jupiter masses (M$_{\rm J}$).  The spectrum paper \citep{1995Sci...270.1478O}, which was in print on 1 December 1995, one day after the discovery paper,  describes the definitive identification of methane and derivation of the temperature in more detail.\footnote{Incidentally, my qualifying exam was scheduled for early November 1995, right in the midst of revising and working to finish the papers and get them accepted for publication.  I decided not to study for the test much, since I thought my work on these two papers should take precedence over an exam.  I do not regret that decision, and, although Shri and I discussed postponing the exam, we decided I should simply do it.  I failed, but I had a really wonderful scientific result.  In the end, I retook the exam and did well, some months later.}

Now and then in science, odd and fascinating juxtapositions of events happen.    Only weeks earlier in June and September, two warm brown dwarfs were announced by \citet{1995AAS...186.6003B} and \citet{1995Natur.377..129R}.  (See Basri's and Rebolo's chapters in this volume.) One week before our result was published, the first definitive detection of a planet-mass object around another star, 51 Pegasi, was also announced \citep{mq95}.  In fact, both were initially announced at the Cool Stars IX meeting in Italy in talks given in rapid succession.  Our work was presented by I. Neill Reid, a close friend and colleague of Shri's who happened to be going to the meeting.

I recall tremendous excitement at the time.  There is nothing like finding something new.  Even if it is not an extremely high-profile result, the euphoria of discovery is a very rare feeling, but one that has kept me working in science.  Actually discovering something new about the universe is partly our duty as human beings, to understand where we are and what this place is made of.  The feeling of having contributed to that process is wonderful, gratifying, and humbling.  Humbling because our crude techniques and attempts to measure light from distant sources is our only way, in astronomy, to uncover what remains hidden in the universe. Yet, it is also incredible how much people have deduced or inferred about the universe in these past two or three centuries.

Over the next few years, I finished the survey of nearby stars, unfortunately without another brown dwarf found.  But I also continued to study Gliese 229B.  As shown in Fig.~\ref{fig:spectrumFull}, through data in the optical, near IR and even out to mid-IR wavelengths of 13~$\mu$m, which our group collected, some fascinating aspects of brown dwarf atmospheres emerged.  In addition, the theorists went to work immediately.  Things such as non-equillibrium chemistry were indicated by the spectrum, perhaps due to huge winds and convection from deeper within the brown dwarf that could cause an apparent imbalance between CO and CH$_4$ (see also the chapter by I.~Baraffe in this volume).  We discovered atomic cesium lines, as well as possible collision-induced absorption due to H$_2$ molecules.  This all was like a playground in which I learned aspects of thermochemistry and atmospheric science.  

Finally, and perhaps most importantly, since astronomers now knew what 1000~K objects looked like, they emerged in greater and greater numbers, especially due to the optical Sloan Digital Sky Survey \citep[and references therein]{2003AJ....126.2081A} which identified the first isolated methane brown dwarf, four years after Gliese 229B was announced \citep{1999ApJ...522L..61S}.\footnote{Dave Golimowski liked to call these ``Sloan Clones'' at the time.} The Sloan discoveries allowed the team working on the infrared 2MASS all sky survey \citep{2006AJ....131.1163S} to refine their search criteria, and many more methane brown dwarfs were immediately identified by both Sloan and 2MASS.  The irony of all this is that the near-IR 2MASS survey, which had a goal of identifying brown dwarfs, could not have found methane dwarfs easily on its own.  Their $JHK$ colors are so close to many stellar objects, such as A-type stars, that optical data combined with IR photometry was necessary.  However, such data was available.

The field has progressed tremendously and various strange types of brown dwarfs are being discovered even now \citep[e.g.][]{2013AJ....145....2F}.  Two new spectral classes, L, which indicates a temperature between M9 and the emergence of methane, and T, which is reserved for objects with methane signatures in their spectra, have been devised \citep{1999ApJ...519..802K,1999AJ....118.2466M,2006ApJ...645.1485B,2008ApJ...678.1372C}.  We have learned that many different processes are involved in determining the emergent spectra of these objects, including age, metallicity, mass, and, perhaps, formation mechanism.  Most recently a third new spectral class may be emerging, called Y, meant to be reserved for objects that show ammonia absorption \citep[][also see Cushing's chapter in this volume]{2011ApJ...743...50C,2012ApJ...753..156K}.  

Some of this new work hints that the brown dwarfs and planets may not be distinct types of objects.  That they may be intrinsically related, or that the two categories have significant overlap in properties.  Time will tell, and perhaps some of the unraveling of the mysterious connection between stars, brown dwarfs and planets will come from the new ability of astronomers to conduct what I like to call ``remote reconnaissance'' of exosolar systems.

\section{Planets and Remote Reconnaissance of Exosolar Systems}
Since 1995, well over 1000 brown dwarfs, a minority of which are companions of stars, and even more planets and planet candidates have been identified with numerous techniques.  As I hope is apparent from the sections above, and the other chapters in this book, the key to understanding these objects in detail is spectroscopy.  However, to observe objects that are millions to billions of times fainter than the stars they orbit and only a fraction of an arcsec away from those stars, very precise control of the starlight is necessary, simply to see them, let alone take spectra of them.  Fortunately there is considerable effort being expended on this problem, by groups in the US, Europe and Japan.  While my own efforts in this direction, Project 1640, are a direct outgrowth of the early' coronagraphy used to find Gliese 229B, the technology involved is considerably more advanced.  It is already revealing new types of objects orbiting nearby stars.  Recently we observed the 1 to 1.8~$\mu$m spectra of all four planets in the HR~8799 system \citep{2013ApJ...768...24O}.  Although there are other instruments similar to Project 1640, such as the Gemini Planet Imager \citep{mgp08}, among others, about to begin observations, I use Project 1640 as an example of what we can do now and what we might expect in the near future for the study of companions of stars.

\begin{figure}[h]
\sidecaption[t]
\centerline{\includegraphics[scale=.47,angle=0]{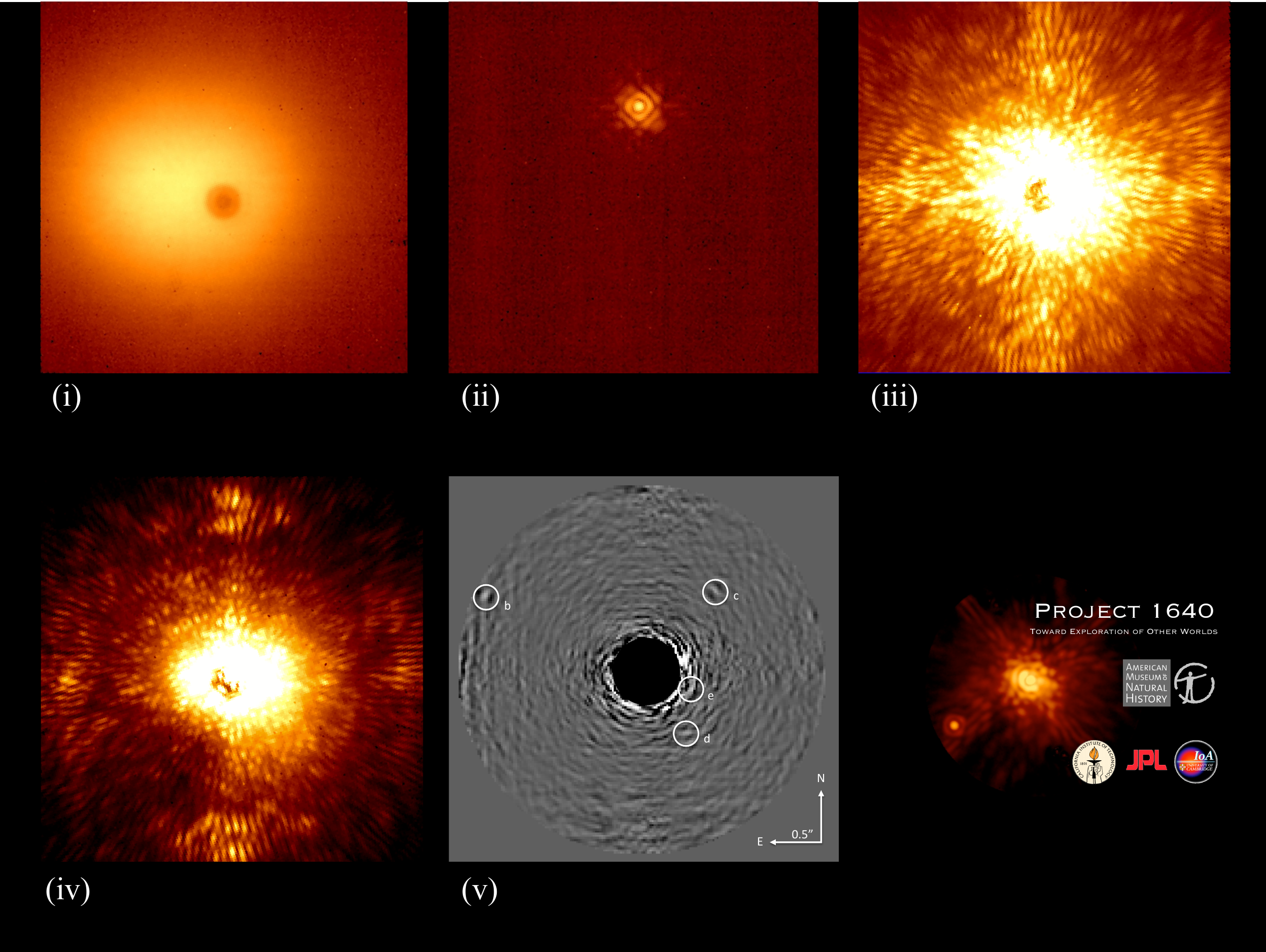}}
\caption{Example of how to conduct a remote reconnaissance of another solar system.  These panels demonstrate how Project 1640 acquires data that can reveal not only a portrait of another solar system, but also spectra of any point sources in that system simultaneously (shown in Fig.~\ref{fig:1640spect}).  (i) The telescope is pointed at the star. The image shows the star before AO is activated. The black spot in the center is the occulting optic in the coronagraph that blocks out starlight. (ii) The AO system has been turned on, to correct the atmospheric turbulence, and the star image is greatly sharpened, reaching the diffraction limit of the telescope. This is a very short 1.5-second exposure. (iii) The star is placed under the occulting optic and a long exposure of five minutes is taken. In this image, most of the starlight has been removed, but a remaining pattern of ÒspecklesÓ fills the field of view.  These are due to defects in the optics. (iv) The calibration wave front sensor is turned on and effects the dimming of the speckles. Numerous long exposures are taken over a 1.25 hour period. (v) The data are assembled and processed with a novel speckle suppression technique based on advances in computer vision to remove the residual starlight and reveal the exoplanets.  Spectra can be extracted once the locations of the planets in the image are determined.  See \citet{obb12} and \citet{2013ApJ...768...24O}. }
\label{fig:1640prog}       % Give a unique label
\end{figure}

Project 1640, an instrument suite involving four separate optical instruments and corresponding control software, with a complex set of custom data reduction and analysis software, is described in \citet{obb12}, \citet{hoz11}, \citet{hob08} and in detail at the level of circuit diagrams, cryogenics, control software, interfaces and opto-mechanical design in \citet{h09}.  The latest system performance metrics are given in \cite{obb12}, including on-sky contrast measurements.  These are described in relation to other projects in high-contrast imaging in \cite{2012SPIE.8442E..04M}, in particular, their figure~1.  In summary, the system is capable of producing images with a speckle floor at roughly 10$^{-5}$ at 1 arcsec separation from a bright star (or 10$^{-7}$ in the lab).  This is achieved through the coordinated operation of four optical instruments: a dual deformable mirror, adaptive optics (AO) system with 3629 actively-controlled actuators, called PALM-3000 \citep[][and 2013, in preparation]{dbr07,dbb06}; an apodized pupil, Lyot coronagraph \citep[APLC;][]{spf09,sl05,s05,saf03eas,saf03}, the design details of which are given in \citet{h09}; a Mach-Zehnder interferometer that senses and calibrates, through feedback to PALM-3000, residual path-length and amplitude errors in the stellar wave front at the coronagraphic occulting spot for optimal diffractive rejection of the primary star's light \citep[CAL;][]{Gautam12,cheng12}; and an integral field spectrograph that takes 32 simultaneous images with a field of view of $3.8 \times 3.8$ arcsec spanning the range $\lambda = 995 - 1769$ nm with a bandwidth of $\Delta\lambda = 24.9$ nm per image \citep[IFS, Fig.~\ref{fig:1640prog}(iii);][]{hoz11,hob08,h09,obb12}.  Aside from technical advances in high-contrast imaging, numerous results from the project include, among others, the discovery and astrometric and spectroscopic characterization of the Alcor AB system \citep{zoh10}, the $\alpha$ Ophiucus system \citep{hmo11}, the $\zeta$  Virginis companion \citep{hob10}, and comprehensive spectral studies of the companion of FU Orionis \citep{2012ApJ...757...57P} and Z CMa \citep{2013ApJ...763L...9H}.

Raw science data generated by Project 1640 are in the form of 2040 $\times$ 2040 pixel images containing 37146 closely packed spectra roughly 30.4 $\times $ 3.2 pixels in extent.  These images are processed into data cubes with dimensions R.A., $\delta$ and $\lambda$, as described in \cite{zbo11}.  

\begin{figure}[t]
\sidecaption[t]
\centerline{\includegraphics[scale=.5,angle=0]{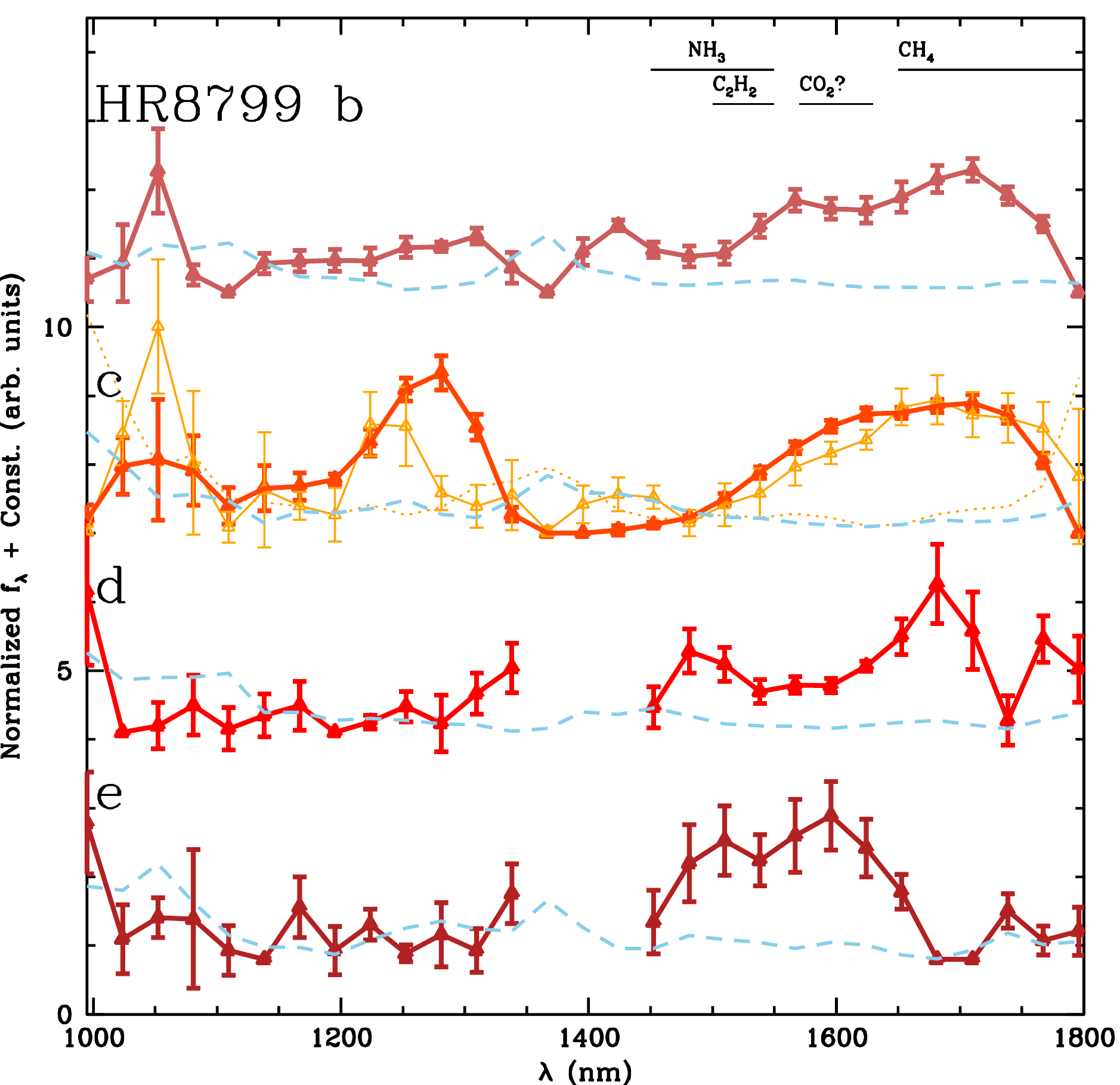}}
\caption{Spectra of the four planets orbiting HR~8799, adopted from \citet{2013ApJ...768...24O}.}
\label{fig:1640spect}       % Give a unique label
\end{figure}

Figure \ref{fig:1640prog} shows how this complex instrument obtains data.  The other projects similar to Project 1640 operate in similar ways, although each has some advantages over the other.  The point is that such a complex system can and does work.  In \citet{2013ApJ...768...24O}, we describe imaging and spectroscopy of the four planets orbiting HR~8799.  This system of four planets was discovered and imaged previously \citep{mmb08,mmv10,mzk10}.  Only limited spectroscopy had been obtained on the two outermost planets \citep{bld10,jbg10}, in addition to a high-resolution $K$-band spectrum of the second most distant planet from the star \citep{2013Sci...339.1398K}.  Project 1640 achieved spectra and images of all four simultaneously with only a little more than one hour of telescope time.   The spectra are shown in Fig.~\ref{fig:1640spect}.  They are all different from each other, and different from any other known objects.  In our initial observational attempt to understand what these planets are, we identified some molecular species that could explain the features in the spectra.  These are entirely tentative and may not bear the scrutiny of proper theoretical models of these objects.  Our preliminary conclusions were the following \citep{2013ApJ...768...24O}:

\begin{itemize}
\item b: contains ammonia and/or acetylene as well as CO$_2$ but little methane.  
\item c: contains ammonia, perhaps some acetylene but neither CO$_2$ nor substantial methane.
\item d: contains acetylene, methane and CO$_2$ but ammonia is not definitively detected.
\item e: contains methane and acetylene but no ammonia or CO$_2$.
\end{itemize}

What are these objects?  They are believed to have effective temperatures around 1000~K, but they do not share many features in common with Gliese 229B or other T-dwarfs, with perhaps the exception of planet e.  Similarly even the lowest gravity L-dwarf spectra do not seem in agreement with any of these objects.

Within the next few years, new campaigns like that of Project 1640 will obtain spectra of, one might hope, several hundred exoplanets, young gas giants, though who knows what we will actually find?  What is clearer than ever is that the population of companions of stars is far more diverse than ever before imagined.  What is the full range of planets and brown dwarfs that may be out there?  Is it meaningful even to use these terms, or do they distract from understanding the real physics of them?

\section{All Known Companions of Stars}

Here, I seek to examine the questions posed above in more detail by examining gross orbital properties of all known companions of stars to date, without binning them into categories people have made (such as star, brown dwarf or planet).  While these properties are not derived from spectroscopy, and they have been measured with many different techniques, my hope, nevertheless, was to find patterns that might reveal that there are, in fact, real groupings or classes of such companions.  My other hope was that perhaps hints of a ``fundamental plane'' or surface or volume within this complex parameter space might emerge, similar to the  fundamental plane for galaxies.

What follows is based on a compilation of orbital properties of some 2413 companions of stars regardless of the mass of the companion or the primary.  Multiple systems are included when available, but systems are only included for which at least M$_1$ (mass of the primary in solar masses, M$_\odot$), M$_2$ (mass of the secondary or higher order component in M$_\odot$), $a$ (physical separation in AU), and $e$ (eccentricity) are known.  Data were collected from published literature and several recent compilations, including \url{exoplanet.eu}, \url{vlmbinaries.org}, the eclipsing binaries reported in \citet{1999ARep...43..521G}, the visual binary orbits in \citet{1999A&A...341..121S}, the 6$^{\rm th}$ orbit catalog of the Washington visual double star catalog \citep[WDS]{2013yCat....102026M}, where companions without magnitude measurements are excluded.  For the WDS companions, {\it Hipparchos}-derived distance is required, and masses were estimated using the mass-luminosity relation of \citet{2009A&A...505..205D} and \citet{2000A&A...364..217D}.  Finally, 102 of the systems listed in \url{bdcompanions.org} with the necessary parameters were included. Datasets for this analysis were last updated in March 2013.

This is an intrinsically biased set of samples, but I did attempt to include as much data as possible. Of note is the fact that Gliese 229B is not included, partly because the orbital parameters are only weakly constrained at this point, with an estimated period of several hundred years.  However, some interesting facts might be hinted at by consolidating all of this data.  Usually researchers tend to look at only subgroups of these categories, based on the notion that stars, brown dwarfs and planets are completely unrelated types of objects.

\begin{figure}[ht]
\sidecaption[t]
\centerline{\includegraphics[scale=.40,angle=0]{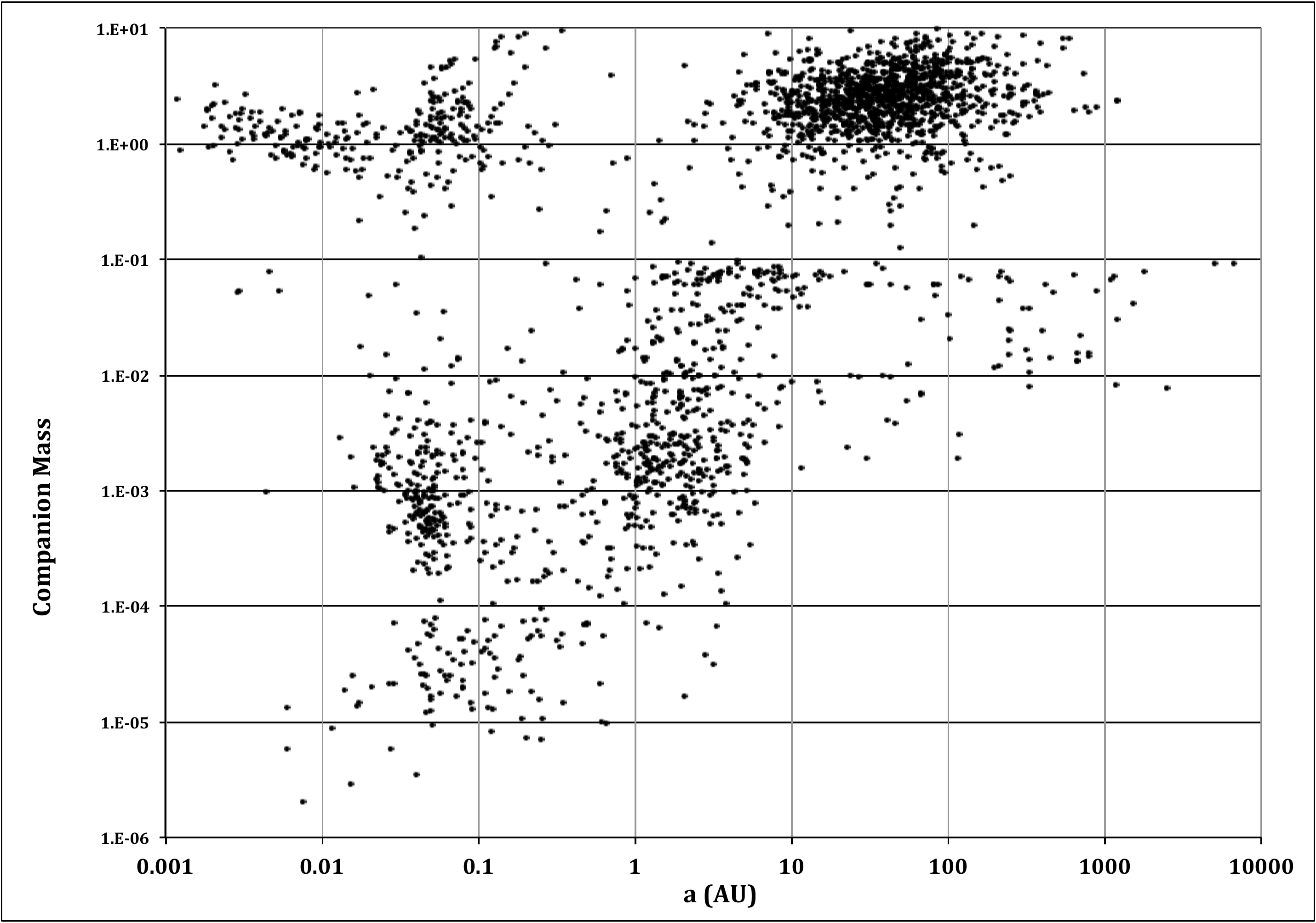}}
\caption{Known companions of stars with orbital elements determined as of March 2013, showing companion mass M$_2$ in solar masses, M$_\odot$, on the ordinate and physical separation in AU on the abscissa.  Data come from a variety of sources, with an attempt to be as thorough as possible with the literature. (See text for data sources.)}
\label{fig:m2va}       % Give a unique label
\end{figure}

In Fig.~\ref{fig:m2va} the data set is represented showing M$_2$ in solar masses  (M$_\odot$) vs.~$a$, a parameter space commonly used in the study of exoplanets.  One might look at this plot and quickly think there are 4 or maybe 5 clumps or groupings of objects: (1) the so-called ``hot jupiters,'' planet-mass objects in very tight orbits below about 0.05 AU; (2) a similarly close clump of stellar mass objects, the eclipsing binaries, below about 0.1 AU; (3) a large clump of higher mass planets centered around 2 AU; (4) a huge agglomeration of stellar companions between 10 and 100 AU; and perhaps (5) a clumping of brown dwarf companions emerging below 0.1 M$_\odot$, from 1 to 10 AU.  

\begin{figure}[h]
\sidecaption
\centerline{\includegraphics[scale=.40,angle=0]{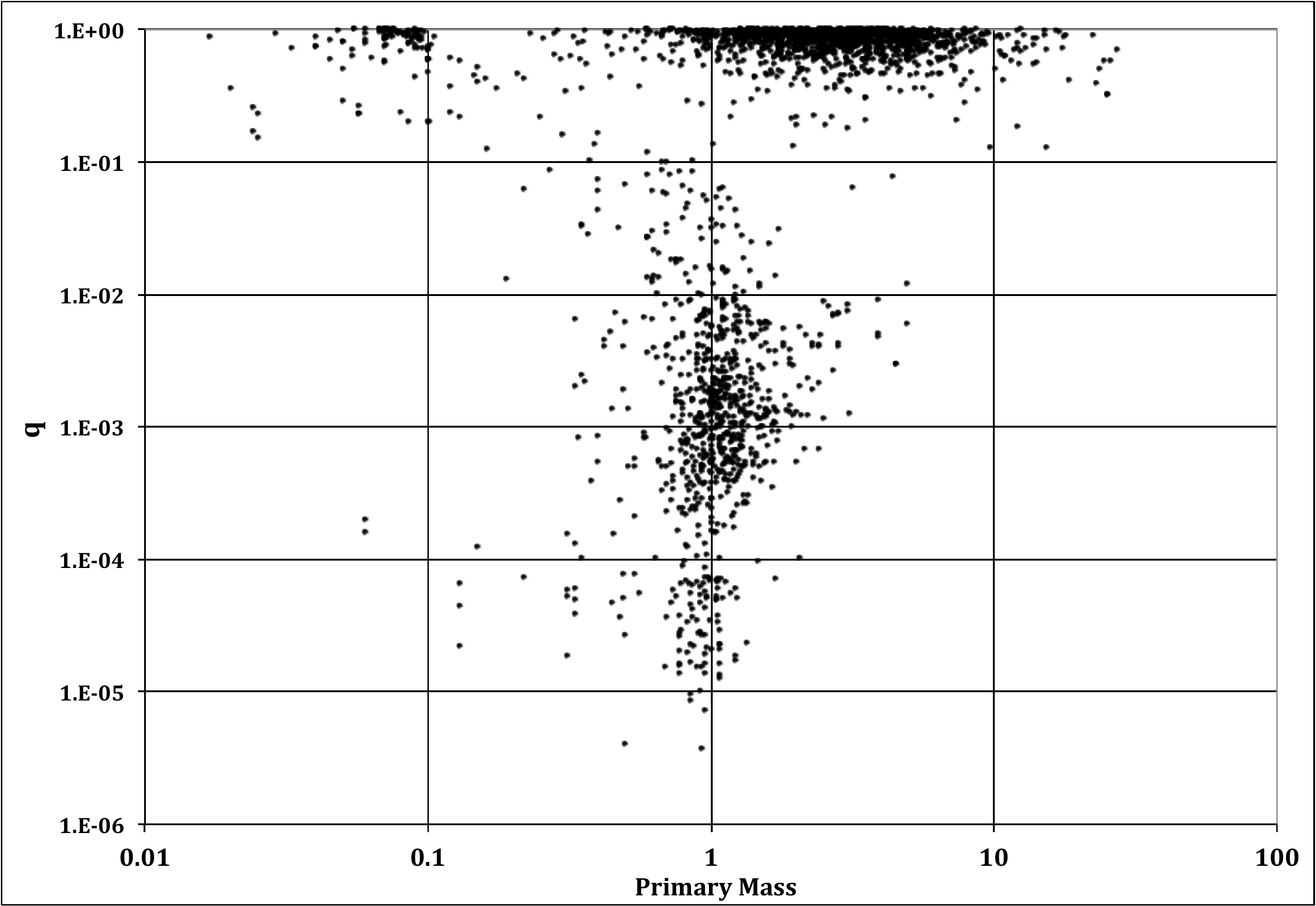}}
\caption{Known companions of stars with orbital elements determined as of March 2013, showing  the mass ratio, $q = $M$_2 / $M$_1$, versus the primary mass in solar masses, M$_\odot$.  Significant biases are present (see text).}
\label{fig:m1vq}       % Give a unique label
\end{figure}

There are several key things to note about this plot.  For years, visual binaries have been subjected to regular study and updated cataloging.  Also for many years, the lower mass companions of stars, especially K and M dwarfs, have been ignored as uninteresting or unimportant.  This results in a dearth of objects in the range from $\sim0.7$ to 0.075 M$_\odot$ (13 M$_{\rm J}$) for M$_2$.  Of course we know that K and M dwarfs make up about 90\% of the population of stars, and in fact many K and M dwarf companions are known with higher mass primaries.  Often these are seen in imaging campaigns and not reported, discarded as uninteresting because a given project may be targeting a different type of companion.  Alternatively, for years some of the radial velocity searches for planets simply stopped monitoring stars that showed radial velocity gradients that were far too large to be due to a planet.  

Another bias is introduced in a similar way by the research on brown dwarfs.  Objects believed to be above about 0.075 M$_\odot$ are ignored as uninteresting.  This is even stranger, because the general, but somewhat unsubstantiated, consensus for a long time has been that brown dwarfs form the same way stars do and that planet formation is a completely separate process.  If brown dwarfs form as stars do, then they should be treated as an extension of the population.  It is possible, from a qualitative inspection of Fig.~\ref{fig:m2va}, that there is an increasing population of brown dwarfs from the lower masses toward the higher, up to that of the lowest mass stars, where the gap in the K and M dwarfs exists.  If that population from the brown dwarfs through to the large population of massive stellar binaries is a continuous single population, that might lend credence to the notion that brown dwarf companions form the same way that binary stars do.  However, it is an equally strong possibility that, as the surveys are increasingly sensitive to these wide brown dwarf companions and planets, the same distribution might also form a continuous grouping with the planets.  The suggestion, considering these biases, then, may be that the companions from 100 AU down to just below 1 AU from the highest stellar masses to well below a jupiter mass constitute a single continuous population.  (n.b. Kumar, in his chapter, suggests that the star formation process has no real lower mass limit.)

\begin{figure}[t]
\sidecaption[t]
\centerline{\includegraphics[scale=.40,angle=0]{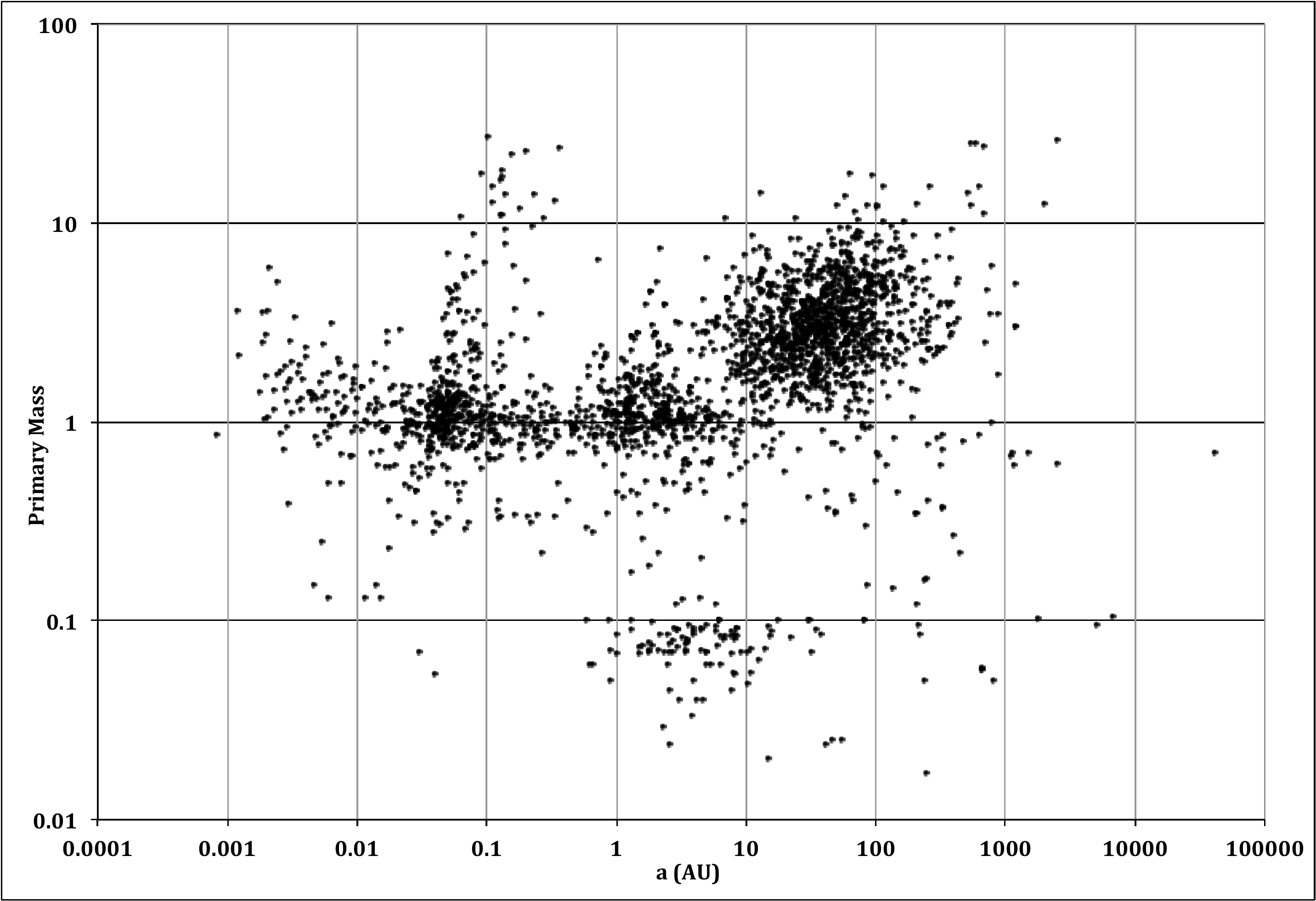}}
\caption{Known companions of stars with orbital elements determined as of March 2013, showing the mass of the primary in solar masses (M$_\odot$) versus the separation of the companion (regardless of mass).  It is unclear that companion separation is a strong function of the type of star a companion orbits.  Note that in this plot most of the planetary systems are represented here near M$_1 = 1 $M$_\odot$.}
\label{fig:m1va}       % Give a unique label
\end{figure}

With regard to the brown dwarf desert---an apparent dearth of brown dwarfs orbiting stars---these plots may shed some light. Shown to be restricted to the range of separations of 0.1 to 1 AU in \citet{oh09}, though still the subject of significant debate, the brown dwarf desert seems apparent in Fig.~\ref{fig:m2va}.  However, there also may be a similar desert of planets in the same separation ($a$) range, and perhaps even among stellar companions.  Is this due to observational bias?  Perhaps only time will tell.  In a fascinating study, though, \citet{2006A&A...446.1165J} suggests that this desert might scale with primary mass, and there may be a hint of this in the data (see also Fig.~\ref{fig:m1va}).  However, from this first plot, it is clear that a full picture will only come from removal of as many observational biases as possible.  The easiest biases to remove are those that are introduced by outright choice in surveys (such as ignoring companions that are not the main target of a particular investigation).

\begin{figure}[t]
\sidecaption[t]
\centerline{\includegraphics[scale=.40,angle=0]{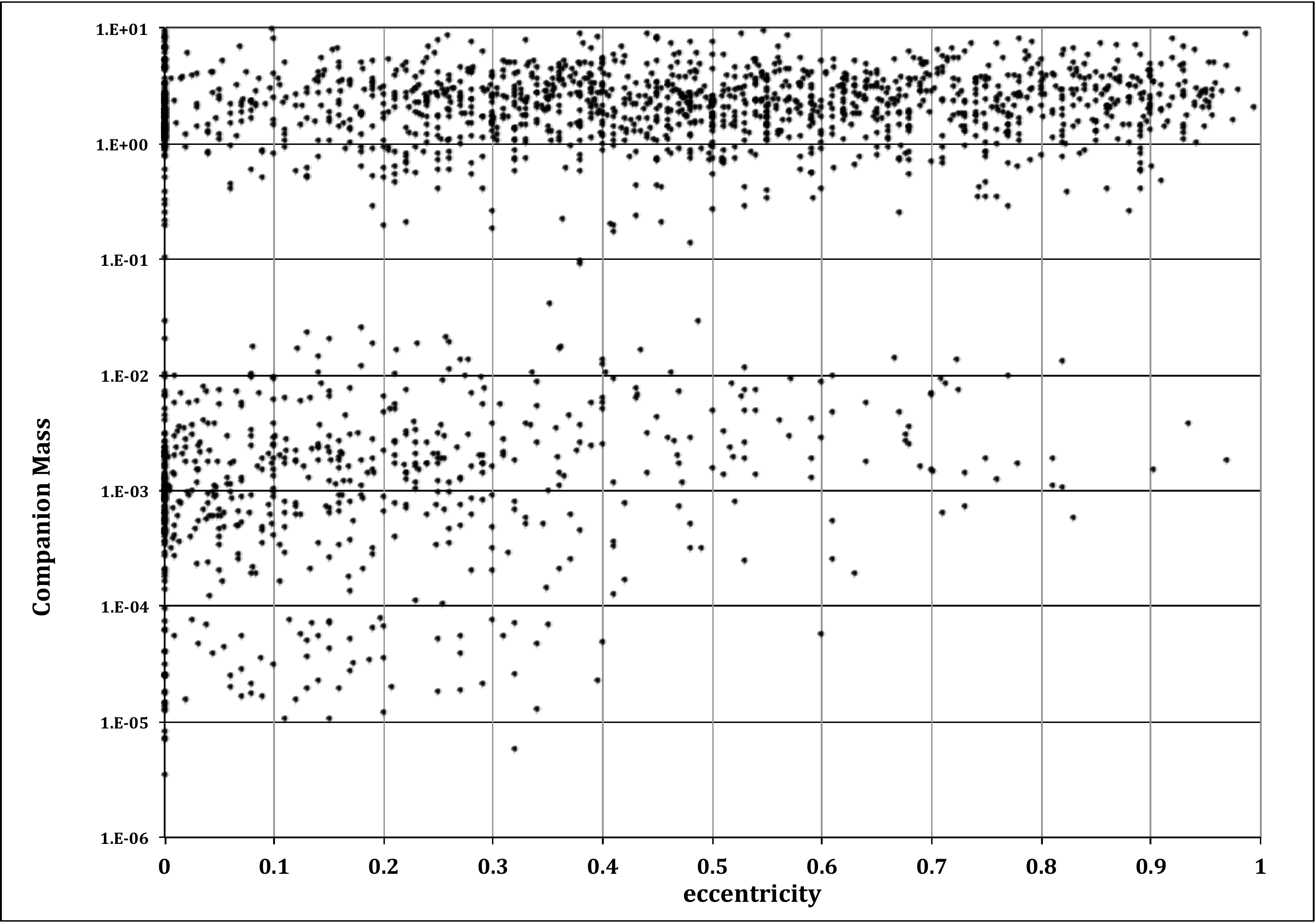}}
\caption{Known companions of stars with orbital elements determined as of March 2013, showing companion mass in solar masses (M$_\odot$) versus eccentricity of the orbit.  The complete lack of objects in the M-dwarf range is likely a heavy observational bias, not a real feature.  However the trend to lower eccentricities in the lowest mass objects may be real.}
\label{fig:m2ve}       % Give a unique label
\end{figure}

The most dramatic depiction of some of the biases in Fig.~\ref{fig:m2va} is shown in Fig.~\ref{fig:m1vq}, where I have plotted the mass ratio, $q$, versus the primary star's mass (M$_1$).  Here one can see that ``nature favors equal mass binaries'' at least for stellar companions.  But again, there is a huge lack of companions of stars below 1 M$_\odot$, due to the bias against studying the most plentiful type of star in the universe, the M- and K-type stars.   Note that a sudden clump arises near the top of the plot below 0.1 M$_\odot$, the product of surveys for binary brown dwarfs.  An even starker picture emerges when looking at the very low-$q$ systems, dominated heavily by the search for exoplanets.  Virtually all exoplanets known are around roughly 1 M$_\odot$ stars.  The new direct imaging campaigns are beginning to find planets around 2 to 5 M$_\odot$ stars, but little is known of planetary systems around M-dwarfs.  In addition, there are a few known planet-mass objects orbiting brown dwarfs.  There is an interesting, perhaps real, lack of companions of sun-like stars between $q = 7 \times 10^{-5}$ and $1.5 \times 10^{-4}$, roughly around 20 to 40 earth masses, intermediate, for reference, between Neptune and Saturn.  Could this be indicative of a different distribution between ice and gas giants, or that perhaps these are two truly different classes of planets?  As a side note, there also seems to be an increasing number of planets toward lower mass, especially below 10 earth masses \citep[e.g.][]{2012NewAR..56...19M}.  Unfortunately these companions are not all represented here due to orbital constraints insufficient to be included in the compilation.

If we extend this investigation further, the previous paragraph suggests that perhaps there may be differences in the types of companions found as a function of the type of primary star.  Thus in Fig.~\ref{fig:m1va} we plot the primary mass vs.~$a$.  Once again the dearth of companions (of all types) within the 0.1 to 1 AU range is obvious.  Even clearer is the observational bias against the low mass stars.  The suggestion that the brown dwarf population is instrinically related to the higher mass stellar primaries is strong here, in the sense that it appears that the clump below 0.1 M$_\odot$ might be a tail of a distribution that is centered at a much higher mass. It is unclear to me that there is a strong function of orbital separations with primary star mass.

As one final example of this companion data, Fig.~\ref{fig:m2ve} shows the companion mass versus eccentricity of its orbit.  This may contain some real information (other than another highlight of bias agains studying M-dwarf companions).  Here we may be seeing a hint that as one proceeds toward the lower mass planets, they tend to favor more circular orbits.  It will be interesting to see how this plot develops as more and more examples are found.

To conclude this section, the study of companions of stars, whether they be brown dwarfs, other stars, or planets, must be conducted broadly, without the bindings of human-chosen categories or nomenclature, in order to arrive at a comprehensive understanding.  There are some suggestions that these three categories are more intrinsically  linked, and that their formation mechanisms may not be entirely distinct, but at least overlapping, or utterly interconnected, modified under certain circumstances due to environment, perhaps.  The study of companions of stars has a long way to go still, but we are at an exciting moment.

\section{Into the Future: Will a companion of a nearby star give humankind its first evidence for biology outside the solar system?}
At this point in history, science has shown that not only are the laws of physics the same in all known parts of the universe, but also that chemistry operates universally.  What we have no evidence for is that there is any biology other than on Earth.  Further, if there is extraterrestrial biological activity, we do not know wether there are any universal aspects to it that are shared with terrestrial biology (such as DNA, or carbon-based biochemistry).  Soon, one may imagine, we may find fish under the ice on Europa, or microorganisms or a skeleton on Mars.  But what about biological activity outside the solar system, as Michell and Herschel clearly showed for gravitational activity over 200 years ago (Sect.~\ref{stat})?

\begin{figure}[ht]
\sidecaption
\centerline{\includegraphics[scale=.6]{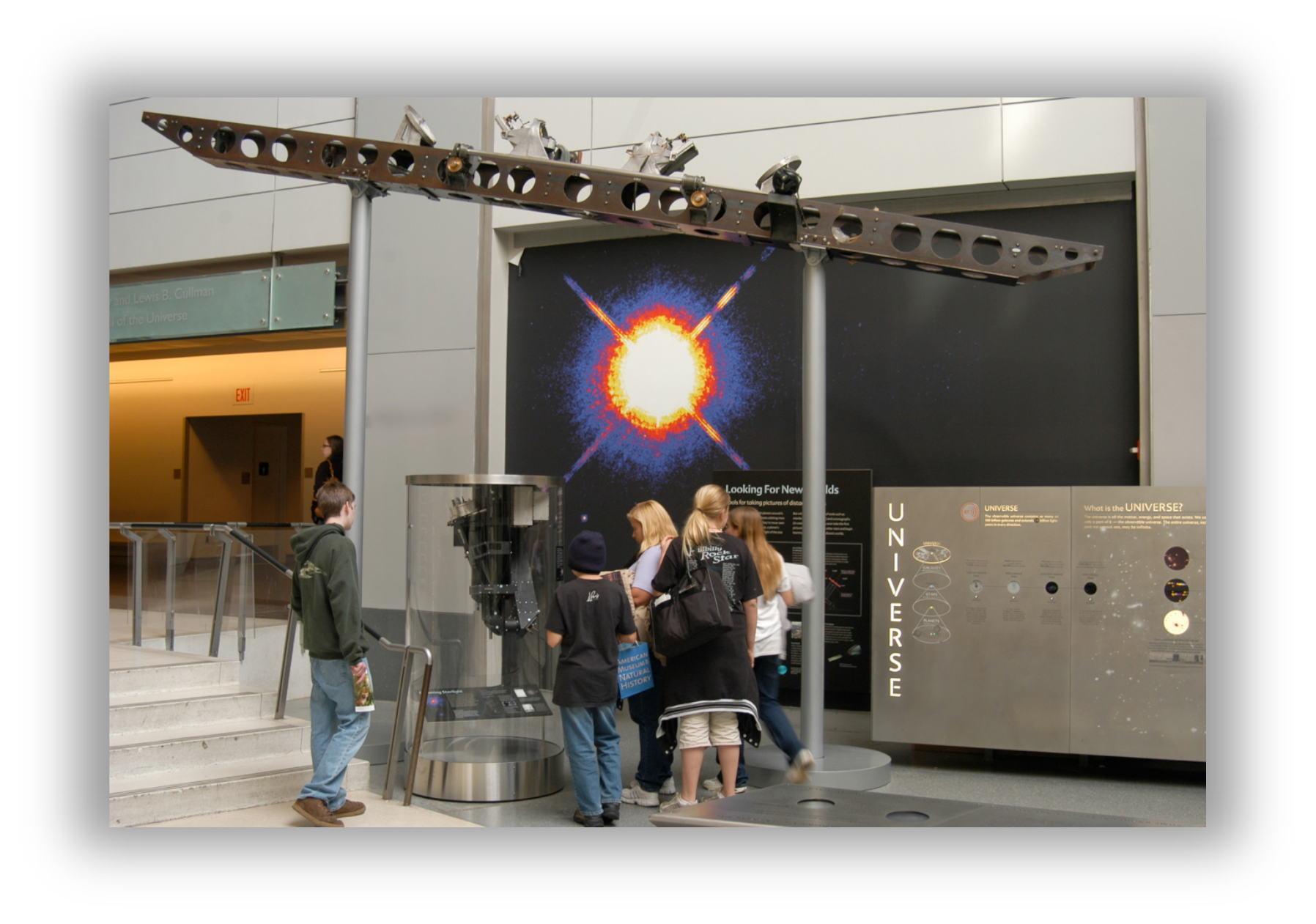}}
\caption{The Adaptive Optics Coronagraph (left-center, inside acrylic cylinder), now on permanent exhibition at the Rose Center for Earth and Space in the American Museum of Natural History.  The exhibit, meant to highlight coronagraphy and interferometry as the principle techniques for direct detection and study of exoplanets, initially included the original interferometer (atop the two columns) that A.~A.~Michelson used on the Mt. Wilson 100-inch telescope to directly measure a number of stellar diameters in the 1920s.  The interferometer is currently on display at Mt. Wilson.}
\label{fig:Exhibit}       % Give a unique label
\end{figure}

I am of the opinion that within the next two decades, at most, such evidence will come to light, and, unless SETI actually discovers complex signals that must be from some form of sentience, the evidence for biological activity outside the solar system will be the direct result of studying the companions of stars.

Such evidence will come with the spectroscopic observation of a planet, one that exhibits an atmospheric composition that cannot be explained by thermochemical equilibrium, or even disequilibrium that is due to strong physical processes in the atmosphere.  Indeed, we can see the evidence for biological activity on Earth in its spectrum, because the abundances of many molecular species are too large without the global forcing and disequilibrium caused by life forms, whether as simple as early bacteria or as complex as people.

I hope that my arguments for studying companions of stars is convincing.  For readers who are interested in details of the state of the art in the field of high-contrast probing of the near vicinities of stars, several recent review articles have been published, including by \citet{oh09}, \citet{am10}, and \citet{2011exop.book..111T}.  This is a field driven by technological advances.  As such, the AOC used to discover Gliese 229B is on permanent display at my home institution, in a small exhibit which exposes the inner workings of the instrument and explains its use (Fig.~\ref{fig:Exhibit}).

\begin{acknowledgement}
The work described regarding Gliese 229B involved many people other than myself and I am deeply grateful for having been able to work with Tadashi Nakajima, Sam Durrance, Shri Kulkarni, Dave Golimowski, Keith Matthews and Marten van Kerkwijk.  It was an honor to conduct such work with them.  Project 1640 also involves a great team of collaborators.  Further information on them and that project can be found at www.amnh.org/project1640.
\end{acknowledgement}

\bibliographystyle{apj} 
\bibliography{MasterBiblio}

\end{document}